\documentclass[journal,draftcls,onecolumn,12pt,twoside]{IEEEtran}
\usepackage{graphicx}
\usepackage{amssymb, amsmath,amsthm, amsfonts}
\usepackage{ifthen}
\usepackage{tikz}
\usepackage{cite}
\usepackage{enumerate}
\usepackage{verbatim}
\usepackage{pifont}
\usepackage{pgfplots}
\usepackage{blkarray}

\usetikzlibrary{arrows}
\usetikzlibrary{patterns}
\allowdisplaybreaks

\newtheorem{theorem}{Theorem}

\newtheorem{lemma}[theorem]{Lemma}
\newtheorem{prop}[theorem]{Proposition}
\newtheorem{defn}[theorem]{Definition}
\newtheorem{remark}[theorem]{Remark}

\newtheorem{claim}[theorem]{Claim}
\newtheorem{example}{Example}

\newcommand{\be}{\begin{equation}}
\newcommand{\ee}{\end{equation}}
\newcommand{\ben}{\begin{equation*}}
\newcommand{\een}{\end{equation*}}
\newcommand{\ba}{\begin{eqnarray}}
\newcommand{\ea}{\end{eqnarray}}

\def\hP{\hat P}
\def\eE{\mathbb E}
\def\bS{\mathcal S}
\def\bP{\mathcal P}

\def\bT{\mathcal T}
\def\eE{\mathbb{E}}
\def\bSl{\mathcal S_{\lambda}}
\def\bbPl{\bar{\mathcal P}_{\lambda}}

\def\bQ{\bar{\mathcal Q}}

\def\bpa{b_1^{\prime}} \def\bppa{b_1^{\prime\prime}}
\def\bpb{b_2^{\prime}} \def\bppb{b_2^{\prime\prime}}

\title{Distributed Scheduling in Multiple Access with Bursty Arrivals under a Maximum 
	Delay Constraint}
\author{Sakshi Kapoor\authorrefmark{2},
	Sreejith Sreekumar\authorrefmark{3},
	Sibi Raj B Pillai\authorrefmark{2}

\thanks{This paper was presented in part at the International Symposium 
on Information Theory, ISIT 2014, Hawai.}
\thanks{\authorrefmark{2} Department of Electrical Engineering, Indian Institute of Technology
Bombay 
email:\{sakshikapoor, bsraj\}@ee.iitb.ac.in}
\thanks{\authorrefmark{3} Department of Electrical and Electronic Engineering, Imperial College London 
email:s.sreekumar15@imperial.ac.uk}
}

\begin{document}

\pgfplotsset{legend style={font=\small,line width=1pt,mark size=1pt},}

\maketitle

\usetikzlibrary{arrows}

\begin{abstract}
A multiple access system  with bursty data arrivals to the terminals is
considered. The users are frame-synchronized, with variable
sized packets independently arriving in each slot at every transmitter. 
Each packet needs to be delivered to a common receiver within a certain number
of slots specified by  a maximum delay constraint. The key assumption  
is that the terminals know only their own packet arrival process, i.e. the  arrivals 
at the rest of the terminals are unknown to each transmitter, except for
their statistics.  For this interesting 
distributed multiple access model,
we design novel online communication schemes which transport 
the arriving  data without any outage,  while ensuring the  delay constraint.
In particular, the transmit powers in each slot are chosen in a distributed 
manner, ensuring at the same time that the joint power vector is sufficient
to support the distributed choice of data-rates employed in that slot. 
The proposed schemes not only are optimal for minimizing the average 
transmit sum-power, but they  also  
considerably outperform  conventional orthogonal multiple access techniques 
like TDMA.
%
%
\end{abstract}

\section{Introduction}
\label{sec:intro}

Multiple access channels (MACs) in wireless systems are conventionally studied under a centralized framework,
where a base-station/controller regulates the transmission rates and powers of
all the users \cite{CovTho91, KnoppHumblet95, TseHanly98, TseHanly98b, TseViswanath05}.
This requires global state knowledge of the underlying time-varying processes.
The lack of such global knowledge in a MAC leads to decentralized operations.
The two common time-varying processes in wireless communication are data-arrivals and
fading coefficients.
Multiaccess under time varying fading models are
extensively studied under centralized frameworks \cite{TseHanly98}, 
decentralized fast-fading setups \cite{ShamaiTelatar99, DasNarayan02, TseViswanath05},
or decentralized block-fading models 
\cite{HwMaGaCi07}, \cite{SDP15jrnl}.
Notice that the fading MACs above assume an infinite bit-pool model, suitable
for mobile applications targeting higher throughputs, without emphasizing
 the delay requirements. As opposed to these, the current paper focuses
on bursty data arrivals to the transmitters, with delay constraints.

Bursty packet arrivals to the terminals are more practical in data networks.
A time-slotted fixed fading MAC with frame-synchronized users and 
independent packet arrivals can effectively
model several limited mobility applications, and wireless back-haul services.
Packets arrive to the respective queue at each transmitter and needs to be
appropriately scheduled through the MAC channel.
It is reasonable to assume here that only the respective transmitters and the receiver
know  the arrival-instants/packet-sizes to each queue~\cite{GamalKim11}. Notice
that bursty arrivals pose new challenges, as it may necessitate data scheduling
and power control to respect the causality of arrivals as well as delay constraints.
While handling arrivals and delays can be challenging in point-to-point channels also,
it is even more pronounced in multiuser networks. More specifically, independent arrival
processes at the terminals of a MAC will force a distributed operation.

The absence of a centralized controller
in a MAC model  will lead to random access. 
However, the name \emph{random access} is
traditionally attributed to dynamic network access schemes like ALOHA, CSMA etc.
These are extensively studied in literature~\cite{RoSi90}. 
In general, the   literature related to network access control falls roughly into
two categories: (i) \emph{closed loop} control and contention resolution; (ii) \emph{open loop}
scheduling and stabilizing queues. ALOHA and CSMA fall into the former 
group, whereas the latter contains flow control schemes based on buffer and 
link states~\cite{GeNeTa06}.  In both models, the objectives typically are to 
maximize throughput, minimize delay, or both.  While the related literature is large, 
in order to  highlight the differences to the model that we consider, 
let us  review some  works relevant to our model.

\subsection{Related Literature}
Closed loop systems like ALOHA and CSMA typically abstract the physical layer as a bit-pipe, where simultaneous
access by several users leads to a collision, or outage~\cite{RoSi90}. Collision events are sensed 
or fed back,  and 
are resolved using contention resolution protocols. While sensing the medium prior to transmission 
can reduce the chances 
of collision,  appropriate control policies are still needed to adjust the transmission
probabilities for achieving optimal throughput~\cite{Bianchi00}. Multi-packet reception capability is also extensively
studied, where it is possible to capture information simultaneously from several users,
see~\cite{CeZuKhMo10} for some recent advances and references. It is well known that 
the  bit-pipe abstraction of physical layer forms an \emph{unconsummated union} with the information theoretic
considerations~\cite{EphreHajek98}.
Several approaches tried to bridge this gap by studying queuing and scheduling models,
by specifying the quality of service constraints by information theoretic quantities
like capacity,
error exponents etc \cite{Telatar95}, \cite{Sibi03}. Under the assumption of reasonably large
blocklengths, these works provide  rigorous mathematical foundations
on which the utilities like transmission-rate and probability of error
can be connected to networking quantities like throughput and delay.

Unlike the statistical multiplexing schemes like ALOHA/CSMA, we consider an
information theoretic MAC model with a fixed
number of users, each observing an independent arrival process. Thus the variability
is not just in the presence or absence of packets, but in the size of the packets itself.
Furthermore, the associated delay constraints may necessitate a packet to be broken into
sub-packets and transmitted in different slots. In this sense, our model differs
from conventional random access. In fact, the model here is more related to 
cross layer scheduling and control in wireless systems, comprehensively
covered in the recent surveys~\cite{Yeh12},~\cite{GeNeTa06}, see also the references therein.
Notice that bursty packet arrivals to  a system  can lead to interesting
trade-offs between the network layer delay and the transmit-power in physical layer, 
and intelligent scheduling algorithms are required to achieve  optimal performance. 
Of particular interest are the \emph{open loop} scheduling schemes which choose the 
transmission parameters such as rate and power based on operating conditions like queue state.

A point to point AWGN link with packet
arrivals was considered in \cite{Rajan04}, with the objective of finding the optimal trade-offs between average
power and delay.
Optimal schedulers which minimize the average transmit power under an average or max-delay constraint
were identified using a dynamic programming (DP) framework.
The key observation in \cite{Rajan04} is that large savings on transmit power can be obtained by
accommodating some  more delay within the tolerable limits.
This was later extended to other scheduling models~\cite{Khoj04}, and also to
networks~\cite{RajanMAC01},~\cite{RajanBC}. Note that all these extensions considered
centralized systems where the arrival processes are known to  all the  terminals.
Interestingly, \cite{RajanMAC01} remarks that the ultimate objective of analyzing centralized
schemes is
to find good \emph{decentralized} schedulers.
We make progress in this direction by
 presenting optimal decentralized
schedulers for a MAC with arrivals, under a maximum delay metric, in the current paper.

In a separate line of work, \cite{PraGamal02} established the optimal energy-efficient
offline  scheduling algorithm which meets a single deadline constraint for
all the arriving packets over a point-to-point AWGN link. The optimal scheduler in this
set up will operate at a low enough transmission-rate, with the
rate at any instant being at least as big as the rates employed till that time. This
leads to the so-called \emph{move-right} algorithm.
An online \emph{lazy} algorithm to vary the transmission rate according to the
current backlog was also proposed and shown to have good asymptotic
performance in \cite{PraGamal02}, see ~\cite{BerryGallager02, Balaji02, TaSuZaMo05, 
Neely06, Neely07, ChNeMi08, LeeJindal09}
for extensions. 
%
%

Energy-delay tradeoffs for multiuser wireless links with online arrivals were considered 
in \cite{Neely07, Neely13}.
In particular, \cite{Neely07} considers a wireless downlink with a separate queue for each
receiver. The base station has global state-information, and 
the broadcast nature of the downlink makes it a centralized model. 
In a more recent work, \cite{Neely13} considers delay aware scheduling
in  multi-user wireless networks. However, a centralized entity schedules one
of the links in each slot.  In contrast to \cite{Rajan04,Khoj04, RajanMAC01, RajanBC, PraGamal02, 
BerryGallager02, Balaji02, TaSuZaMo05, Neely06, Neely07, ChNeMi08, LeeJindal09, Neely13}, which all had
some form of centralized scheduling and control, a decentralized MAC with arrivals is 
considered in this paper. 

Models with both time-variations in arrivals and fading coefficients are
also of interest. For example,  ~\cite{BerryGallager02, Balaji02, FuMoTs06}
consider dynamic  fading and arrivals  for a point-to-point system, whereas
\cite{Neely06, Neely07, Neely13} analyze centralized multi-user models.
In another interesting work, \cite{QiBe04} considers  a slow-fading distributed MAC,
 where each user has access only to its own link quality and arrival process,
from a collision resolution perspective. Along the same lines, \cite{QiBe06}
proposes a channel aware ALOHA protocol to exploit multiuser diversity.
A centralized scheduler with decentralized power control is considered 
for contention resolution in \cite{ElEp04}. Notice that 
\cite{QiBe04, QiBe06, ElEp04}  do not explicitly
address any  delay  constraints, and employ the underlying
physical layer bit-pipe view of random access. Taking a different standpoint, 
efficient decentralized open-loop schedulers for a fading MAC with arrivals,
so as to minimize the average sum-power required to
communicate in  an outage-free manner, is an interesting problem. 
To keep the average power bounded, one can assume that the possible fading values of interest
are non-zero. 
This is one of the topics discussed in this paper, for which there seems
few prior results. 

Perhaps the closest work in literature to the current sequel is the distributed rate-adaptation 
framework in a block-fading MAC~\cite{SDP15jrnl},
and its application to energy harvesting~\cite{KP16}. However, both \cite{SDP15jrnl} and \cite{KP16}
consider throughput maximization in distributed MACs, and have nothing to do with
delay constraints. More specifically, \cite{SDP15jrnl} maximizes the throughput under local knowledge
of the link fading parameters, whereas \cite{KP16} achieves the same objective under the distributed knowledge
of energy harvesting processes at the transmitters. Interestingly, one of the motivations behind  
the introduction of  a distributed rate-adaptation framework in \cite{HwMaGaCi07} was the 
throughput maximization in random  access systems.
\emph{Broadcasting} is another useful technique to increase the throughput of
distributed systems, where depending on the conditions,  parts of the data can be 
correctly decoded~\cite{MiFrTs12}. Rate-less coding without any arrivals for 
distributed multiple access was considered in~\cite{Niesen06}.
As opposed to these, the objective of the current paper is in 
minimizing average sum-power under a maximum delay constraint. This, in some sense, 
parallels the problem of throughput maximization in distributed systems \cite{SDP15jrnl, KP16}.
In fact, the approach and techniques here are motivated by \cite{SDP15jrnl, KP16}, this will be
evident from the structural similarities of the results presented here.

\subsection{Main Contributions}
We consider a  $L-$user AWGN MAC with bursty packet arrivals, as shown in Figure~\ref{fig:one}. 
The transmissions are frame-synchronized, and time is divided into slots or blocks (the words
`slots' and `blocks' are used interchangeably in this sequel). 
Assume that variable sized packets
independently arrive  at the respective terminals at the start of each slot. 
The packets are to be conveyed to the receiver within $D_{max}$ slots, i.e. a max-delay constraint
of $D_{max}$.
Each transmitter, by observing  its own data arrival stream, 
will schedule the transmission rate as well as power in a slot-wise manner such that 
the arrived data is conveyed before the respective delay constraints. 
The challenge here is to perform successful  data transfer without knowing
the exact arrivals 
at the other terminals, except for the statistics. 
The word \emph{successful} is used in
the sense of transmitted data not being in \emph{outage} for any transmission block.  
Notice that no arrival in a slot is also allowed, it is considered as a zero sized packet. 
We consider transmission  schemes which will not only guarantee
successful communication, but also  minimize the 
average transmit sum-power expenditure. In short, we seek power efficient communication
schemes for a distributed MAC with online arrivals.

Notice that  we assumed the observation of independent random processes at 
different transmitters.  The techniques here depend crucially on the knowledge  
at each terminal of the statistics of all time-varying quantities in the system. 
The MAC receiver is also aware of the realizations of all the random variables in each
slot.  
 The statistics are only used in the initial design phase,
the proposed communication schemes will still work even if the underlying
statistics are perturbed. However, the optimality guarantees do not hold
under perturbations.
In other words, once the statistics are conveyed, no further information exchange
is necessary for designing the distributed communication scheme.

%

The main contributions of the current paper are:
\begin{enumerate}
\item An optimal  distributed communication scheme for a MAC with independent
bursty data arrivals is presented under a unit slot delay constraint on the arriving packets. 
An explicit power allocation scheme is shown to give an almost closed form solution 
to the minimal average transmit  sum-power (Theorem~\ref{thm:two}, Section~\ref{sec:fix:fade}).
\item 
An  optimal distributed power control policy incorporating both time-varying fading 
and bursty arrivals is presented, for a unit slot delay constraint (Theorem~\ref{thm:optpwr}, 
Section~\ref{sec:var:fade}).
\item  
 For a general max-delay constraint of $D_{max}$, and a fixed fading  MAC with 
independent bursty arrivals, 
we propose  an iterative technique to find optimal schedulers for rate-adaptation 
and power control (Section~\ref{sec:del:dmax}). 
This effectively addresses the question posed in \cite{RajanMAC01}: ``the 
ultimate goal is
to find decentralized schedulers that approach the performance
of the centralized scheduler".
\end{enumerate}
Our results capture the tradeoff between the QoS parameters of delay
and required energy/power, for a distributed wireless multiple access model in which
several users can simultaneously access the medium. Notice that the users
are free to do rate adaptation  and power control, while ensuring outage free
operations.  The  trend of tolerable delay being proportional 
to the achieved energy  efficiency is an expected one, this is observed in the
distributed MAC model too. However, the results clearly
demonstrate that higher energy efficiency and lower delay than conventional schemes
can be simultaneously achieved by resorting to the optimal  communication
schemes presented in this paper.

The techniques here also apply to more general delay constraints than max-delay. However,
max-delay is chosen for its simplicity as well as wide application. 
In particular,
the proposed communication schemes can be extended to other delay constraints for which  efficient
single user schedulers can be identified. Also, the utility of average sum-power is chosen 
for convenience,
the results can be extended to minimize the weighted average sum-power as well.


The paper is organized as follows. Section~\ref{sec:model} details 
the system model and notations. Section~\ref{sec:fix:fade} considers distributed MACs with
fixed fading values and bursty arrivals, under a unit slot delay constraint. 
In Section~\ref{sec:var:fade}, we extend the unit slot delay results to  
the case of  dynamically varying fading and bursty arrivals. Then, in  Section~\ref{sec:del:dmax}, 
we consider a  fixed fading MAC under
a general max-delay constraint of $D_{max}$ slots, and propose  an iterative algorithm
to compute the optimal average sum-power in this case.
Simulation results are provided in Section~\ref{sec:sim:1}, Section~\ref{sec:sim:2}
and Section~\ref{sec:sim:3}, to compare the performance of the optimal schemes proposed
here with the conventional  schemes in literature. Finally, 
Section~\ref{sec:conc} concludes the paper.

In this paper $\eE[X]$ denotes  the expectation of random variable $X$. 

\section{System Model} \label{sec:model}
\begin{figure}[htbp]
\begin{center}
\includegraphics[scale=0.75]{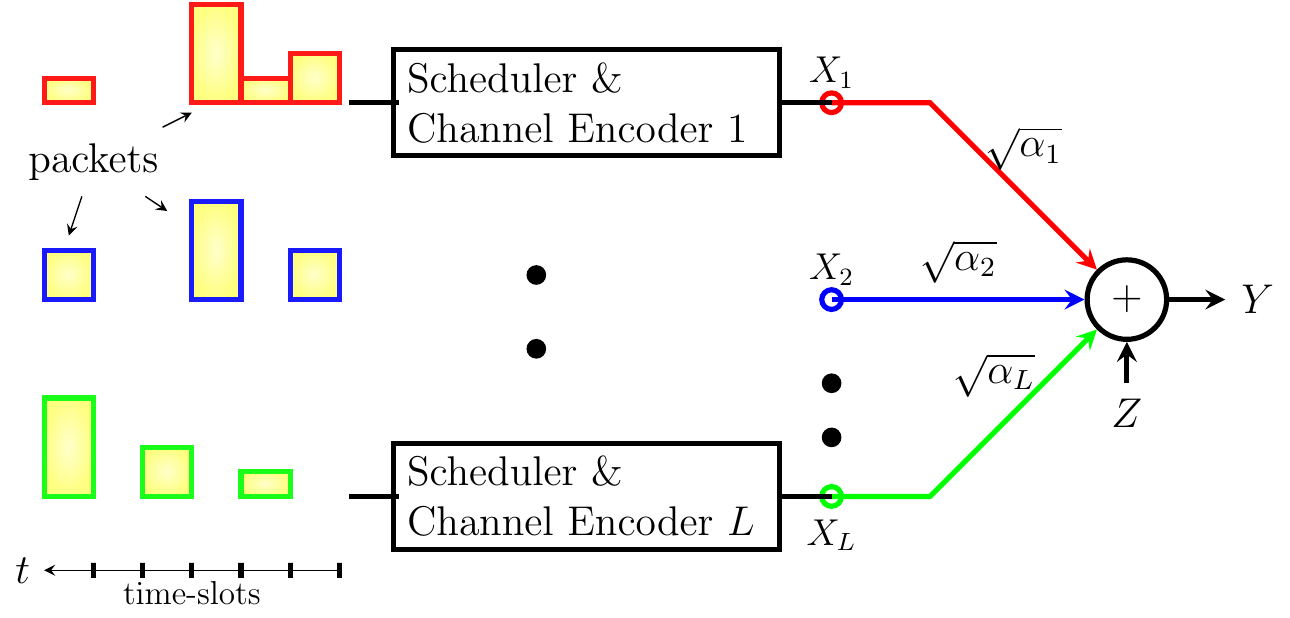}
\vspace*{-0.25cm}
\end{center}
\caption{Distributed MAC with bursty packet arrivals \label{fig:one}}
\end{figure}

 Consider the multiple access system shown in Figure~\ref{fig:one}, which is referred to
as a
{distributed MAC with bursty arrivals}.  For $L$ transmitters,
 the real valued discrete-time model is described by the observed samples
$$
Y = \sum_{i=1}^L \sqrt{\alpha_i} X_i + Z,
$$
where $X_i$ represents the transmitted symbols from user~$i$. The fading coefficients
$\sqrt{\alpha_i}, 1 \leq i \leq L$ are  assumed to be fixed and known to all parties.  
The noise process $Z$ is normalized  additive white Gaussian,
independent of all the transmitted symbols.  The transmissions take place in a
frame-synchronized slotted manner, where each slot (or block) is of length $N$. The blocklength $N$ is
assumed to be large enough for coding and decoding to take place with a sufficiently low 
error probability.

At the start of each time slot, a variable sized packet arrives independently at each transmitter.
We denote the arrival process to terminal~$i$ as $A_i[j]$, which implies that $NA_i[j]$ bits
arrive at the start of block~$j$ to this terminal. 
The most important aspect of the system that we consider is 
that each transmitter knows only  its own arrival process, i.e.
the packet-sizes at rest of the terminals are unknown to each transmitter. However, the statistics
of all the arrival processes are available  to each party.  For simplicity as well as practical relevance, 
we will assume that $A_i[j]$ are independent and  identical across~$j$, each taking
values from a finite set $\mathcal A$, with $|\mathcal A| < \infty$.
Furthermore, we also assume that the arrivals at different terminals are independent, but can be of
arbitrary distributions on $\mathcal A$.

\begin{figure}[htbp]
\begin{center}
\begin{tikzpicture}[yscale=0.7,xscale=0.85]
\draw[->] (0,0) --++(0,5) node[above]{$P_2$};
\draw[->] (0,0) --++(5,0) node[below]{$P_1$};
\draw[line width=1.5pt] (1,4.5) -- (1,2) -- (2,1) --(4.5,1);
\draw[dashed,|-|] (1,2) --++(0,-2) circle (0.025cm)node[below]{$\frac{2^{2r_1}-1}{\alpha_1}$};
\draw[dashed,|-|] (2,1) --++(-2,0) circle (0.025cm) node[left]{$\frac{2^{2r_2}-1}{\alpha_2}$};
\path[pattern=dots] (1,4.5) -- (1,2) -- (2,1) --(4.5,1) --++(0,3.5) -- cycle;
\draw[dotted,left to-right to,thin] (3,0) node[below]{$\frac{2^{2(r_1 + r_2)} - 1}{\alpha_1}$} -- (0,3)node[left]
{$\frac{2^{2(r_1 + r_2)} - 1}{\alpha_2}$};
\end{tikzpicture}
\caption{Set of $(P_1,P_2)$ supporting rate-pair $(r_1,r_2)$ \label{fig:contra} }
\end{center}
\end{figure}
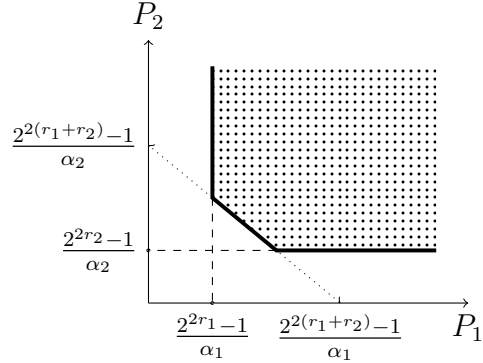

Assume that  each packet is required to be delivered within 
$D_{max}$ time slots of its arrival.
In the system model depicted in Figure~\ref{fig:one}, each transmitter is shown to have two
components, a scheduler and  a channel encoder.  The scheduler specifies the number of bits
to be conveyed in each slot, or equivalently, the transmission rate. 
Notice that the system allows multi-slot breakup of packets without
violating the max-delay of each packet.  The channel encoder has to ensure that the scheduled
bits in each slot are conveyed correctly to the receiver, i.e. there is no outage. More precisely,
we say that the receiver does not encounter outage if the decoding error probability
in each block decays exponentially to zero with blocklength, a standard practice in
information theory parlance~\cite{GamalKim11}, see \cite{Sibi03} for a more formal justification. 
It is well known that any rate-tuple inside
the AWGN MAC capacity region will not lead to  outage in the above sense. 
Thus, for a rate-vector $(r_1, \cdots, r_L)$ in a block, 
the channel encoders can ensure successful decoding by choosing Gaussian codebooks with
high enough short-term (or per-slot) average transmit power $P_i$ at terminal $i\in\{1, \cdots, L\}$ such that
\begin{align} \label{eq:pow:contra}
\sum_{i \in J} \alpha_i P_i  \geq 2^{2(\sum_{i\in J} r_i)} - 1 , \forall J \subseteq \{1,\cdots, L\}.
\end{align}
Thus, for any rate-vector $(r_1,\cdots,r_L)$ scheduled in a slot, the transmit powers
should obey \eqref{eq:pow:contra}.
For a two user
MAC model, the set of power-tuples which can support a rate-pair $(r_1,r_2)$ is illustrated in
Figure~\ref{fig:contra} as
the shaded portion, which is a contra-polymatroid~\cite{TseHanly98}. 
%
%

\begin{figure}[htbp]
\begin{center}
\includegraphics[height=5cm,width=12cm]{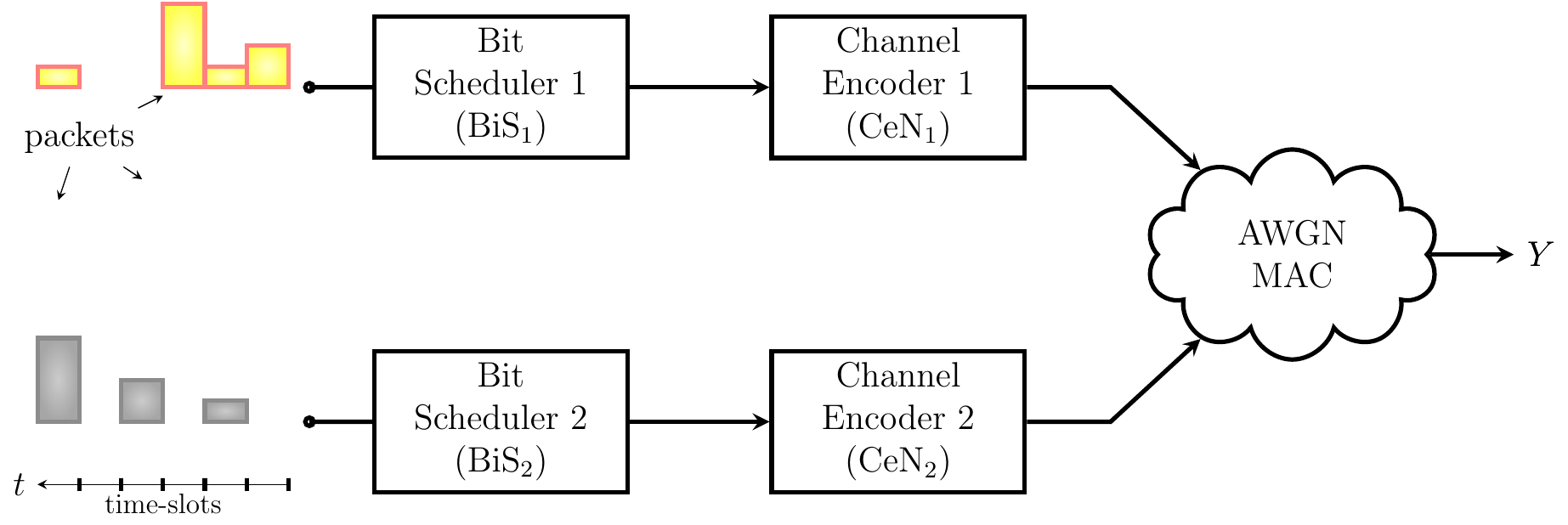}
\end{center}
\caption{Decoupling of Transmitters into BiS and CeN~\label{fig:tandem}}
\end{figure}

\begin{defn} \label{def:outage:free}
A set of power allocation functions $P_i(\cdot), 1 \leq i \leq L$
satisfying \eqref{eq:pow:contra} for any feasible
rate-vector $(r_1, \cdots, r_L)$ is called an outage free power allocation.
\end{defn}
We consider only outage free power allocations in this paper. In addition,
each terminal has to do rate-adaptation,  which specifies
the number of bits scheduled for transmission in a slot-wise manner, while ensuring
the maximal delay constraint. Schemes meeting the max-delay constraint with
outage free power allocations are called as \emph{outage free communication schemes}. 
%
Since the exact arrivals as well as rate-demands at other terminals are not available, 
each transmitter makes scheduling decisions based on  its own arrival history, along with
the statistics of arrival processes at all the terminals.  
Let $N B_i[j]$ bits are scheduled for slot $j$ by terminal~$i$. 
In other words, $B_i[j] \in \mathcal B_i$ is the transmission rate chosen for slot~$j$ at user~$i$. 
The remaining bits will wait in the queue for future scheduling. 
At the start of block~$j$, let $N.\hat r_i[j,d]$ be the number of bits remaining in the 
$i^{\text{th}}$ queue 
which can afford a delay of at most $d$ more blocks. Note that $\hat r_i[j,D_{max}] = A_i[j]$.
\begin{defn}\label{def:state:vec}
The  $D_{max}-$dimensional
vector $\zeta_i[j] = \bigl(\hat r_i[j,d],\, 1\leq d \leq D_{max}\bigr)$ is termed as 
the state-vector of transmitter~$i$.
\end{defn}
At times we may drop the square brackets and call
 the state-vector as $\zeta_i$. 
Our objective is to compute the infinite horizon minimum average sum-power expenditure 
$P_{avg}^{min}(D_{max})$
at the terminals, i.e.
\begin{align} \label{eq:emp:power}
P_{avg}^{min}(D_{max}) := \inf_{\Theta}\limsup_{M\rightarrow \infty}  \sum_{l=1}^L \frac 1M 
	\eE \left( \sum_{j=0}^{M-1} P_l(B_l[j]) \right),
\end{align}
where $\Theta$ is the set of all \emph{outage free} communication schemes which specify the 
rate-power tuples $(B_l[j],P_l(B_l[j]))$, $1\leq l \leq L, 0 \leq j \leq M-1$, while 
meeting the maximal delay constraint $D_{max}$ for each packet.  The formulation in 
\eqref{eq:emp:power} is actually the infinite horizon average cost minimization problem of
a Markov Decision Process (MDP)~\cite{Bertsekas2001}, \cite{Sennott99}. Such MDPs 
already find wide applications in single user scheduling problems~\cite{Rajan04}. 
In the MDP formulation, the scheduling actions at terminal~$l$ are based on the 
current value of $\zeta_l$, i.e. the size and delay requirements of the queue backlog.
For $1\leq l \leq L$, let
$\Theta^d_l$ be the collection of all deterministic outage free strategies 
$\theta_l: \zeta_l \mapsto (B_l,P_l)$,  with $(B_l,P_l) \in \mathcal B_l \times \mathbb R^+ \bigcup \{0\}$, 
such that the packet-delay at user~$l$ is at most $D_{max}$ for any $\theta_l \in \Theta_l^d$. 
Observe that no queue in the system ever builds up, since we have
bounded packet-sizes and a maximal delay constraint.
Furthermore, in the AWGN MAC setup that we consider, 
it is also reasonable to assume that the per block average power 
at a transmitter is continuous in the transmission-rate.
These observations allow the following reformulation of~\eqref{eq:emp:power}.
\begin{lemma} \label{lem:reform}
\begin{align}\label{eq:det:power}
P_{avg}^{min} (D_{max}) =  \sum_{l=1}^L  \inf_{\theta_l \in \Theta^d_l}\lim_{M\rightarrow \infty}  \frac 1M \eE \sum_{j=0}^{M-1} P_l(B_l[j]).
\end{align}
\end{lemma}
\begin{IEEEproof}
The proof is given in Appendix~\ref{sec:app:reform}.
\end{IEEEproof}
Under the reformulation in Lemma~\ref{lem:reform}, notice that $B_l[j]$ can be taken as the output process of
a deterministic scheduler with IID arrivals as inputs. Thus $B_l[j]$ is a stationary
ergodic process and we can write~\cite{Bertsekas2001} 
$$
P_{avg}^{min} (D_{max}) =  \sum_{l=1}^L \inf_{\theta_l} \eE \left( P_l( B_l) \right),
$$
where the random variable $B_l$ has distribution same as the marginal ergodic law of $B_l[j]$.
Now we can focus on 
designing optimal power allocation schemes using the distributions of
$B_l, 1 \leq l \leq L$. This  effectively decouples each transmitter into two components, 
\emph{viz.} a bit scheduler (BiS) and a channel encoder (CeN).
%
%
This is illustrated  in Figure~\ref{fig:tandem} 
for a two user MAC model.

Each bit-scheduler (BiS) ensures that the delay constraint $D_{max}$ of every arriving 
packet is met.  In addition to meeting the delay constraint, the BiS works in tandem 
with the channel encoder (CeN) to improve the overall power efficiency. 
On the other hand,  each CeN operates under a unit delay constraint,  
ensuring that the bits scheduled by the BiS for every slot are 
successfully conveyed to the receiver by the end of that slot.
%

Let the set of $L$ BiSs and CeNs employed at the transmitters be 
denoted by $\bar{\bS}$ and $\bar{\bP}$ respectively, 
we will use $S_i$ to refer to BiS~$i$, and $P_k$ for the power law of CeN~$k$. 
Thus $S_i:\zeta_i[j] \mapsto B_i[j]$, where
 $B_i[j] \in \mathcal B_i$ is the transmission rate chosen for slot~$j$ at user~$i$. 
When the context is clear, we call $\bar \bS$ as the scheduling scheme, and $(\bar \bS, \bar \bP)$
as the communication scheme.
The following example illustrates the scheduler actions for a two user MAC.
\begin{example}
A pair of schedulers  with $\mathcal A = \{1,2,3\}$ and $D_{max}=2$ 
is shown in Figure~\ref{fig:mat1},  where the row and column indexes
respectively indicate the elements of the two dimensional state-vector $\zeta_i$. 
The matrix entries specify the scheduled transmission-rate for that
state-vector. For example, from state $(1,2)$ at the start of block~$j$ for user~$1$, 
a transmission-rate of $2$ will be chosen. Then, the new state-vector at the start of block~$j+1$ is $(1,A_1[j+1])$,
where the second entry can withstand a delay of $2$ units. 

 \begin{figure}[htbp]
 \[ 
S_1 =\begin{blockarray}{cccc}
&  1 & 2 & 3  \\
\begin{block}{c [ccc]}
  0 & 1 & 2 & 2 \\
  1 & 2 & 2 & 2 \\
  2 & 2 & 2 & 2 \\
  3 & 3 & 3 & 3\\
  \end{block}
\end{blockarray} \qquad
S_2 =\begin{blockarray}{cccc}
&  1 & 2 & 3  \\
\begin{block}{c [ccc]}
  0 & 1 & 2 & 2 \\
  1 & 2 & 2 & 2 \\
  2 & 2 & 2 & 2 \\
  3 & 3 & 3 & 3\\
  \end{block}
\end{blockarray}
\]
\caption{Schedulers $S_1$ and $S_2$ for $\mathcal A=\{1,2,3\}$, $D_{max}=2$}\label{fig:mat1}
\end{figure}
\end{example}
The schedulers shown in Figure~\ref{fig:mat1} output integer-valued transmission rates. However,
we can in general allow real-valued rates to be chosen. 
In practice, the schedulers maybe limited to choose rates which are  multiples of some 
small quanta, or pick one from a given finite set of rates. The effect of quantization on scheduled
rates will  be illustrated further in the numerical studies of Section~\ref{sec:sim:3}.

While the techniques proposed in this sequel extend to any AWGN MAC with  independent
bursty arrivals, for simplicity, we demonstrate most of the results for a two user MAC.
Let the respective fading coefficients be $\sqrt{\alpha_1}$ and $\sqrt{\alpha_2}$, 
with $\alpha_1 \geq \alpha_2$. %
In order to proceed with the optimization of \eqref{eq:det:power},  we first define a notion of
time-sharing between two communication schemes $(\bar \bS, \bar \bP) $  and 
$(\bar{\bT}, {\bQ})$.
While this notion is useful for our proofs, we reiterate that the optimal distributed
schemes in this paper do not employ time division multiple access (TDMA). In fact, the proposed 
schemes can considerably outperform any variant of TDMA based communication schemes.

\subsection{Time sharing of Scheduling Schemes} \label{sec:time:share}
The \emph{time sharing} that we introduce here is a bit different from the conventional
time division scheme, the latter has different schedulers employed in non-overlapping
time intervals. On the other hand, a conceptual  time sharing is used here to construct 
a new scheduler  from two existing schedulers, and both schemes will have an 
impact in each slot of data transfer.
In particular, two BiSs are combined to simultaneously operate on the online arrivals
as follows.  
\begin{defn}
Consider two scheduling schemes $\bar\bS$ and 
$\bar{\bT}$, both meeting a maximal delay of $D_{max}$. 
For $k= S,T$ and $l=1,2$, let $B_{k_l}[j]$
denote the rate scheduled in slot~$j$ by user~$l$ under the 
scheduling discipline~$k$, when the same arrival process is fed to the
two schedulers.
For $\lambda \in (0,1)$, define a 
new scheduler $\bar{\bS}_{\lambda}$ such that user~$l$
schedules  a rate $\lambda B_{S_l}[j] + (1-\lambda)  B_{T_l}[j]$
for slot~$j$.  
\end{defn}
\begin{lemma}
The scheduler $\bar{\bS}_{\lambda}$ is a valid scheduler meeting 
the maximal delay constraint of $D_{max}$.
\end{lemma}
\begin{IEEEproof}
Suppose each packet from an arrival process is split into two with a fraction
$\lambda$ of the bits going to the first segment. Let us add dummy bits to each of these segments
to make their sizes same as that of the original packet. Thus we obtain two identical
streams of data, and can apply $\bar\bS$ and $\bar\bT$ separately on these. Since both $\bar\bS$
and $\bar\bT$ meet the delay constraint, we have shown that a fraction $\lambda$ of
the bits get routed through $\bar\bS$, and the remaining through~$\bar\bT$. Offloading the
dummy bits and combining the remaining streams will give us $\bar\bS_{\lambda}$.
\end{IEEEproof}
Let us also define a time-sharing on the power-allocation functions. 
Let $P_{avg}(\bar \bS, \bar \bP)$ be the average sum-power for the communication scheme 
$(\bar \bS, \bar \bP)$.
\begin{defn}
Consider two power allocations $\bar \bP$ and $\bQ$, which allocate 
powers $(P_1(b_1),P_2(b_2))$ and $(Q_1(b_1),Q_2(b_2))$ respectively to support
a rate-pair of $(b_1,b_2)$. The \emph{time-shared power allocation} $\bbPl$ allocates
$\bigl( \lambda P_1(b_1) + (1-\lambda) Q_1(b_1), \lambda P_2(b_2) +(1-\lambda) Q_2(b_2) \bigr)$ for 
 $(b_1,b_2)$.
\end{defn}

\begin{lemma}\label{lem:convex:sp}
Consider two communication schemes $(\bar\bS,\bar\bP)$ and $(\bar\bT,\bQ)$, and their time-sharing $(\bar\bSl,\bbPl)$.
Then, $(\bar\bSl,\bbPl)$ is an outage-free communication scheme and
\begin{align} \label{eq:convex:sp}
P_{avg}(\bar\bSl, \bbPl) = \lambda  P_{avg} (\bar\bS,\bar\bP) + (1-\lambda) P_{avg}(\bar\bT, \bQ).
\end{align}
\end{lemma}
\begin{IEEEproof}
The lemma essentially means that the average sum-power $P_{avg} (\bar\bS,\bar\bP)$ 
is convex in the pair $(\bar\bS,\bar\bP)$.
Let us choose any possible scheduled rate-pair $(\bpa,\bpb)$ from $\bar\bS$. Since $\bar \bP$
can successfully support this rate-pair, the 
corresponding received power obeys
$$
\alpha_1 P_1(\bpa) +  \alpha_2 P_2(\bpb)  \geq 2^{2(\bpa + \bpb)} - 1.
$$ 
Similarly for a rate-pair $(\bppa,\bppb)$ from $\bar\bT$, we have
$$
\alpha_1 Q_1(\bppa) +  \alpha_2 Q_2(\bppb)  \geq 2^{2(\bppa + \bppb)} - 1.
$$
However,
\begin{align}
\lambda .  (2^{2(\bpa + \bpb)} - 1) + (1-\lambda). (2^{2(\bppa + \bppb)} - 1) 
	\geq 2^{2(\lambda (\bpa + \bpb) + (1-\lambda)(\bppa + \bppb))} - 1,
\end{align}
by the convexity of the function $2^{x}$ for $x \geq 0$.  Thus,
\begin{align}
\alpha_1 \left( \lambda P_1(\bpa) \!+\! (1-\lambda)Q_1(\bppa) \right)
	+ \alpha_2 \left( \lambda P_2(\bpb) \!+\! (1-\lambda)Q_2(\bppb) \right)	
	\geq 2^{2(\lambda (\bpa + \bpb) + (1-\lambda)(\bppa + \bppb))} - 1.
\end{align}
This guarantees that the scheme $\bbPl$
can support every rate-pair scheduled by $\bar\bSl$. 
Thus $(\bar\bSl,\bbPl)$ is an outage-free communication scheme. Furthermore,
the  average sum-power of  $(\bar\bSl,\bbPl)$ is same as the $\lambda-$
linear combination of the average sum-powers individually achieved by $(\bar\bS,\bar\bP)$ and 
$(\bar\bT,\bQ)$ respectively, completing the proof.
\end{IEEEproof}
We now present optimal scheduling schemes for our distributed MAC model. 
The next two sections discuss the case of unit slot delay constraint, i.e. $D_{max}=1$.
\section{Optimal Power Adaptation Under a Unit Delay Constraint} \label{sec:fix:fade}
Consider the  system shown in Figure~\ref{fig:tandem} with the
BiS as an identity function, i.e. all remaining bits are scheduled for transmission
at the start of each block, yielding $A_l[j] = B_l[j], \forall j$. This will
correspond to a unit slot delay constraint~\cite{SPD14c}.
The arrivals are assumed to be IID over slots, but they have independent, otherwise arbitrary,
distributions across users. 
The IID assumption is for simplicity,  the results easily generalize to 
stationary ergodic processes at the terminals.
We will  first propose a  lower bound to the average sum-power expenditure, 
and then construct a scheme which meets this bound. The approach here can be visualized
as a dual to the MAC throughput maximization framework of \cite{SDP15jrnl}. However
\cite{SDP15jrnl} does not consider arrivals or delay constraints, 
rather, throughput maximization under a distributed
CSIT assumption in time-varying fading models is pursued.

Let the bit-rate random variable $B_i$ at terminal~$i\in\{1,2\}$ be discrete with the marginal law 
\begin{align} \label{eq:lam:1}
Pr(B_i = b_{ik})  = \lambda_{ik} , 1 \leq k \leq K_i,
\end{align}
where the values $b_{ik}$ are assumed to be increasing in $k$, and $K_i$ is the
cardinality of the support of $B_i$.
The CDF of $B_i$ is represented by $\phi_i(b)$. 
In order to properly combine different integrals, we define an inverse CDF function $b_i(x), i=1,2$ 
for $x\in [0,1]$, given by
\begin{align} \label{eq:inv:cdf}
b_i(x) = \phi_i^{-1}(x) := 
	\begin{cases} \sup\{b \in \mathbb{R} |\, \phi_i(b) < x \} \text{ for } 0 < x \leq 1 \\
		\sup\{b \in \mathbb{R} |\, \phi_i(b) \leq  x \} \text{ when }  x=0.
	\end{cases}
\end{align}
%
Using \eqref{eq:inv:cdf}, and by a change of variables
\begin{align} 
\eE[P_i(B_i)] = \int_{\mathbb R^+} P_i(b) d\phi_i(b) 
	  = \int_0^1 P_i(b_i(x))dx.  \label{eq:expect:inv:2}
\end{align}
Notice that the integral 
expression shown in terms of the CDF works even when the underlying distribution is discrete
as  $b_i(x)$ is defined for all $x\in[0,1]$.
We can now express our result in terms of $b_i(x)$. 

\begin{theorem} \label{thm:two}
For a two user MAC with independent bursty arrivals, 
and respective fading coefficients of  $\sqrt{\alpha_1}$ and $\sqrt{\alpha_2}, \,  \alpha_1 \geq \alpha_2$,
the minimum average sum-power required under a unit slot delay constraint $D_{max}=1$ is
$$
P_{avg}^{min}(1) = \int_0^{1-\frac{\alpha_2}{\alpha_1}} \frac{2^{2b_2(x)} - 1}{\alpha_2} dx + 
	 \int_0^{\frac{\alpha_2}{\alpha_1}} \frac{2^{2(b_2(v+1- \frac{\alpha_2}{\alpha_1}) 
	+ b_1(\frac {\alpha_1 v}{\alpha_2}))}-1}{\alpha_2} dv.
$$
\end{theorem}

\begin{IEEEproof} 
Though the expression above
appears complex, the minimum sum-power expenditure is simple to evaluate for any set
of independent arrival processes.
The proof proceeds by starting with the expectation expression in \eqref{eq:expect:inv:2} and
constructing a suitable lower bound as $x$ traverses from $0$ to $1$. This is given in
the coming subsection. An outage-free communication
scheme operating at this average sum-power will then be presented in \ref{sec:achieve}, thus proving the theorem. 
\end{IEEEproof}

\subsection{Lower Bound to $P_{avg}^{min}(1)$} \label{sec:lower:bnd}
Let us denote $P_i(b_i(x))$ as $\hat P_i (x)$, $\frac{\alpha_2}{\alpha_1}$  as $\alpha$,
and take $\bar \alpha = (1-\alpha)$.
The expected sum-power can be written as
\begin{align}
\eE[P_1(B_1) + P_2(B_2)] \!\!\!\!\!\! \hspace*{-2.0cm}& \notag\\ 
 &= \int_0^1 P_1(b_1(x)) + P_2(b_2(x)) dx \notag \\
 &= \int_0^{\bar\alpha} \hP_2(x) dx + \int_{\bar\alpha}^1 \hP_2(x)dx + \int_0^1 \hP_1(x) dx \notag \\
 &= \int_0^{\bar\alpha} \hP_2(x) dx  + 
	\int_{0}^{\alpha} \left( \hP_2(v+ 1- \alpha) + \frac{\hP_1(\frac{v}{\alpha})}{\alpha} \right) dv
	\label{eq:ser:1}\\
 &\geq \int_0^{\bar\alpha} \frac{2^{2b_2(x)} - 1}{\alpha_2} dx + 
	\int_{0}^{\alpha}  \frac{\alpha_2 \hP_2(v+ 1- \alpha) + \alpha_1 \hP_1(\frac{v}{\alpha})}{\alpha_2} dv \label{eq:ser:2}\\
 &\geq \int_0^{\bar\alpha} \frac{2^{2b_2(x)} - 1}{\alpha_2} dx + 
	 \int_0^{\alpha} \frac{2^{2(b_2(v+\bar \alpha) + b_1(\frac v{\alpha}))}-1}{\alpha_2} dv \label{eq:ser:3}.
\end{align}
In the above, \eqref{eq:ser:1} is obtained by change of variables and combining two integral terms. The inequality
\eqref{eq:ser:2} results from the fact that an average power of 
$\alpha_2^{-1} \left( 2^{2b} - 1\right)$ is required
to transmit at a rate of  $b$ bits per transmission by user~$2$, even when the other user is absent.
Furthermore, to support  the rate-pair $(b_1,b_2)$, we know from  \eqref{eq:pow:contra} that
%
\begin{align}
\alpha_1 P_1 + \alpha_2 P_2 \geq 2^{2(b_1 + b_2)} - 1,
\end{align}
which will in turn justify  \eqref{eq:ser:3}.  Thus our converse proof is complete.

\subsection{Scheme achieving $P_{avg}^{min}(1)$} \label{sec:achieve}
We will specify an iterative outage free communication scheme with  an average power of 
$P_{avg}^{min}(1)$ given in Theorem~\ref{thm:two}.
Notice that it is sufficient to specify the corresponding transmit 
power against the rates given by $b_i(x), 0 \leq x \leq 1$,
these are the inverse CDF values defined in \eqref{eq:inv:cdf}.


%
Let us denote $\frac{\alpha_2}{\alpha_1} = \alpha$, and $\bar{\alpha} = 1 - \alpha$.
Motivated by \eqref{eq:ser:3}, we can assign
\begin{align} \label{eq:pow:2:init}
P_2(b_2(x)) = \frac{2^{2b_2(x)} - 1}{\alpha_2}\,,\,\, 0 \leq x \leq  1-\alpha,
\end{align}
to match the first term there.  The rest of the allocations are chosen 
to match the remaining terms in \eqref{eq:ser:3}.
%
%
%
To this end, define
$$
m = \max\{k: \sum_{i=1}^{k-1} \lambda_{2i} < \bar \alpha \},
$$ 
where $\lambda_{2i}$ is given in \eqref{eq:lam:1}.
Now, consider the set 
\begin{align} \label{eq:gamma:def}
\Gamma_{\text{no}} := \{0\} \bigcup \bigl\{\sum_{i=1}^j \!\lambda_{2i} - \bar \alpha,\, m \leq j \leq K_2 \bigr\} 
\bigcup \bigl\{\sum_{i=1}^j \alpha\lambda_{1i}, 1 \leq j \leq K_1 \bigr\}.
\end{align}
Let us arrange the elements of $\Gamma_{\text{no}}$ in  ascending order to obtain an ordered set $\Gamma$. 
Observe that the set $\Gamma:=\{\gamma_0, \gamma_1, \cdots, \gamma_{|\Gamma|-1}\}$ 
includes all the CDF values of $B_1$ scaled by a factor $\alpha$, in addition to
other terms. Thus
the set $\{b_1(\frac{\gamma_k}{\alpha}), \forall k\} =  \{b_{1k}, \forall k\}$, 
where $b_{1k}$ is the $k^{th}$ biggest bit-rate required at user~$1$.
Similarly  
$\{b_2(\gamma_k + \bar \alpha), \forall k \} = \{b_{2k}, k \geq m\}$. 
%
The power allocations are  iteratively specified for the corresponding values in 
the increasing order of $\gamma_i$.
After each assignment, the iterative procedure computes the power for
a hitherto unassigned bit-rate value, chosen based on the ordered list $\Gamma$.
By convention, user~$2$ is updated before the other whenever possible.  
%
Using the short notation, 
$$
P^{\mathrm s}_{u,v} :=2^{2\left(b_2(u+1-\frac{\alpha_2}{\alpha_1})+b_1(\frac{v \alpha_1}{\alpha_2})\right)}-1,
$$
we are all set to specify the power allocations. 
%
\begin{defn} \label{defn:opt:pow}
Let $P_1(\cdot)$ and $P_2(\cdot)$ be two power allocation functions such that
 \begin{align} \label{eq:lem:pow:1}
P_2(b_2(x)) &= \frac{2^{2b_2(x)} - 1}{\alpha_2} ,\, 0 \leq x \leq 1 - \frac{\alpha_2}{\alpha_1} 
\end{align}
and for $\gamma_i \in \Gamma,\, 0 \leq i \leq  |\Gamma|-1$,
\begin{align}
\alpha_1 P_1\left(b_1(\frac{\gamma_i \alpha_1}{\alpha_2})\right) &= P^{\mathrm s}_{\gamma_i, \gamma_i} 
	- \alpha_2 P_2\left(b_2(\gamma_{i} + 1-\frac{\alpha_2}{\alpha_1})\right) 
        \label{eq:lem:pow:2} \\
\alpha_2 P_2\left(b_2(\gamma_{i+1} +1-\frac{\alpha_2}{\alpha_1} )\right) 
	&= P^{\mathrm s}_{\gamma_{i+1}, \gamma_{i}}  - \alpha_1 P_1\left(b_1(\frac{\gamma_{i}\alpha_1}{\alpha_2})\right) .
        \label{eq:lem:pow:3} 
\end{align}
Recall that $\Gamma$ is the set given in \eqref{eq:gamma:def} arranged in the ascending order.
\end{defn}

\begin{lemma} \label{lem:achieve:1} The power allocations given in 
\eqref{eq:lem:pow:1} -- \eqref{eq:lem:pow:3} achieve $P_{avg}^{min}(1)$ over a two user 
distributed MAC with bursty arrivals. 
\end{lemma}
\begin{IEEEproof}
It is clear that the transmit powers can be chosen as mentioned in the lemma. 
On close observation of our achievable scheme, we have matched the terms given in 
the derivation of the lower bound in  Section~\ref{sec:lower:bnd} with equality. 
This will guarantee that our scheme indeed has the minimum
possible average power expenditure over a distributed MAC with bursty arrivals and a unit delay constraint. 
The only missing part is to show that every transmission rate-pair corresponding to the incoming packets can 
be sustained without outage by the chosen power allocation. This is 
proved in the next section for the more general case of bursty arrivals as well as dynamic fading, see 
Lemma~\ref{lem:outage:2}. The proof of Lemma~\ref{lem:achieve:1} is now complete.
%
\end{IEEEproof}




\begin{remark} \label{rem:erg:ext}
The proof of Lemma~\ref{lem:achieve:1} can be adapted  to continuous-valued distributions on the arrivals $A_i[j], i=1,2$,  
and also to 
arbitrary stationary ergodic arrival processes which are independent across the terminals. The
former case is detailed in Appendix~\ref{sec:cntvalpkts}.
\end{remark}

\subsection{Simulation Study} \label{sec:sim:1}
Let us now study a simple example  to  show the utility of the proposed results. Consider
a two user MAC system with fading coefficients $1$ and $\sqrt{\alpha}$ respectively.
Let the required bit-rate in a slot be chosen from $\{1,2\}$  and the arrival law at each terminal be 
based on independent and identical Bernoulli random variables with 
$Pr(B_i=1)=0.75, i=1,2$. 
Let us first compare the sum-power of our scheme with two TDM-based schemes.
In simple TDM (S-TDM),  
users share each slot equally among them, whereas in generalized TDM (G-TDM), the fraction of time allotted to 
a user is optimized to minimize the total transmit power.

\begin{figure}[htbp]
\begin{center}
\includegraphics[scale=0.65]{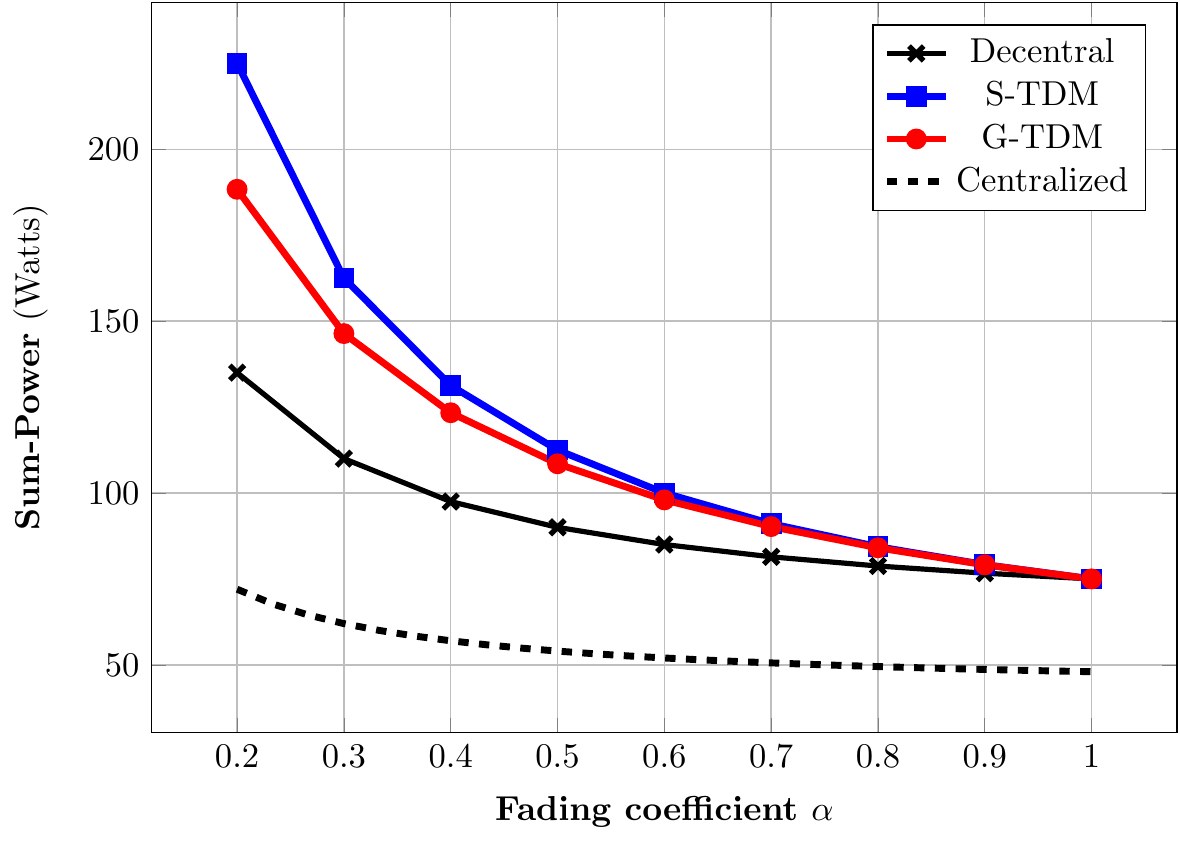}
\caption{Optimal Vs TDM for two user distributed MAC with $\alpha_1=1, \alpha_2=\alpha, D_{max}=1$. \label{fig:comp:energy}}
\end{center}
\end{figure}

Figure~\ref{fig:comp:energy} compares the power expenditure when the link parameter $\alpha$
is varied in $[0.2,1]$. The average sum-power for the optimal decentralized scheme
is shown as `Decentral'.
When $\alpha$ moves away from $1$, it is evident that there is considerable
advantage in using the proposed optimal scheme, over alternatives like TDMA. For a lower bound,
we have also plotted the average sum-power of an optimal centralized scheme (Centralized), where each
 terminal  has the global knowledge of arrivals at all the users.

\subsection{Structural Properties of Decentralized Power Allocation} \label{sec:unit:slot:struct}
Before generalizing the optimal decentralized schemes, let us highlight some procedural and
 structural aspects of the optimal decentralized power allocation, the latter are used in the coming sections.

Observe that each terminal has access to the causal knowledge of its own arrival process, along with
the statistics at all the terminals. Before the start of any communication, each user can compute its 
power allocation as a function of the rate requirement. This is done using Definition~\ref{defn:opt:pow}, which only
relies on the global statistics.  For communicating, the pre-computed power allocation is used
to map each arrived rate in a slot to a corresponding transmit power, and a corresponding codeword. 
This only requires individual knowledge of the
arrivals at each terminal. Remarkably, the distributed choice of powers never leads to outage in
any block. In other words, the chosen power tuple can sustain the arrived rate vector requirement,
as the resulting MAC capacity region is guaranteed to contain the operating rate-pair. In addition,
the scheme also minimizes the average sum-power consumption, thus making it optimal.

Let us now list some structural aspects.

\begin{lemma}\label{lem:conv:rate}
Each of  the power allocation functions $P_i(\cdot),i=1,2$ given in Definition~\ref{defn:opt:pow} 
is  convex in the rate.
\end{lemma}
\begin{IEEEproof}
The proof is given in Appendix~\ref{sec:app1}.
\end{IEEEproof}
Notice further that though the power-allocations in 
Lemma~\ref{lem:achieve:1} are given for a set of
rates specified by \eqref{eq:inv:cdf}, the iterations can be continued to 
extrapolate for higher rate-values, if desired. This can be done by
adding suitable dummy rates of zero probability.
In addition, one can also extend each allocation to any continuous interval of rates by 
time-sharing. Lemma~\ref{lem:convex:sp} guarantees that the resulting communication scheme
is outage free.  We summarize these observations as a remark.
\begin{remark} \label{rem:rate:pow}
Using the power allocation scheme in Lemma~\ref{lem:achieve:1}, we can
define a single user scheduler with rate-power characteristics $P_l(b), b\in [0,|\mathcal B_l|]$
at terminal~$l$, using time-sharing and extrapolation.
\end{remark}
See Figure~\ref{fig:pow:conv} for an illustration of the rate-power characteristics. Let
us now incorporate dynamic fading to our model.

\def\cHi{|\mathcal H_i|}
\def\cHa{|\mathcal H_1|}
\def\cHb{|\mathcal H_2|}
\def\cBi{|\mathcal B_i|}
\def\cBa{|\mathcal B_1|}
\def\cBb{|\mathcal B_2|}
\section{Dynamic Channels and Bursty Arrivals}\label{sec:var:fade}

Consider a scalar two user  discrete-time AWGN MAC with independent bursty arrivals, where the
channel coefficients also vary independently across links. Each user knows  its own  
transmission-rate requirement  as well as its  fading coefficient at the start of the block.
Let the arrivals to terminal~$i$ be IID with the required rate distribution
$Pr(B_i=b_{ik})= p_{ik}$. 
The channel $H_i$ undergoes independent block fading with  $Pr(H_i=h_{ik})=q_{ik}$. 
We assume a finite number of positive fading values for each link in our  MAC model. 
Let us arrange $b_{ik}$ and $h_{ik}$ such that  they
are increasing in $k$ for each~$i$.
For $i=1,2$, let $\phi_i$ be the CDF of the arrival process $B_i$,
and $\psi_i$ be the CDF of $H_i$. The objective is to find the power allocation schemes 
$P_i(b_{ij},h_{ik}), i=1,2$ which minimize the average sum-power, i.e.
\begin{align}
P_{avg}^{min}(1) = \min_{P_1(\cdot, \cdot),P_2(\cdot, \cdot)} \eE_{\phi_1,\psi_1}  \left( P_1(B_1,H_1)\right)+ \eE_{\phi_2,\psi_2}\left(P_2(B_2,H_2) \right) \label{avgsmpwr} .
\end{align}
Recall that $P_i(\cdot, \cdot)$ only depends on $(B_i, H_i)$ due to the
distributed system assumptions.
Let $\cBi$ and $\cHi$ denote the cardinality of the sample space of $B_i$ and $H_i$ respectively.
Define $\alpha_{0\cHa}=0$ and $\beta_{0\cHb}=0$,  and let
\begin{align} \label{eq:def:level}
 \alpha_{jk}=\alpha_{(j-1)\cHa}+\sum_{l=1}^k\frac{p_{1j}q_{1l}}{h_{1l}^2},~ 1 \leq j \leq \cBa,~ 1 \leq k \leq \cHa \notag \\
\beta_{jk}=\beta_{(j-1)\cHb}+\sum_{l=1}^k\frac{p_{2j}q_{2l}}{h_{2l}^2},~ 1 \leq j \leq \cBb,~ 1 \leq k \leq \cHb.
\end{align}
Let us illustrate these definitions and notations by an example.
\begin{example} \label{eg:2}
Take  $B_1\in\{2,3\}, H_1\in\{1,\sqrt 3\}, B_2\in\{1,2\}, H_2\in\{1,\sqrt 2\}$, with
$Pr(B_1=2) = \frac 13$,  
$Pr(H_1=1) = \frac 14, 
Pr(B_2=1) = \frac 14, 
Pr(H_2=1) = \frac 12$.
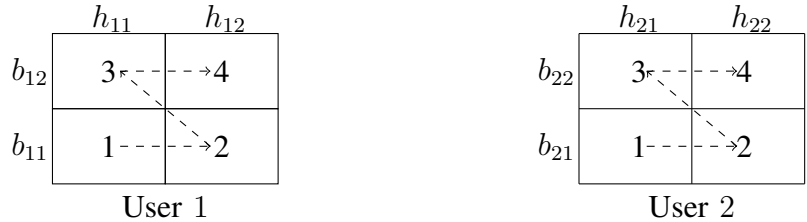
\begin{figure}[htbp]
\begin{center}
  \begin{tikzpicture}
  \draw (0,0)rectangle (1.5,1)node[midway] {3};
  \draw (1.5,0)rectangle (3,1)node[midway] {4};
  \draw (0,0)rectangle (1.5,-1)node[midway] {1};
  \draw (1.5,0)rectangle (3,-1)node[midway] {2};
  \node at (0.8,1.2) {$h_{11}$};
  \node at (2.3,1.2) {$h_{12}$};
  \node at (-0.3,0.5) {$b_{12}$};
  \node at (-0.3,-0.5) {$b_{11}$};
  \node at (1.5,-1.3) {User~$1$};
  \draw [dashed,->] (0.9, -0.5) -- (2.1, -0.5) ;
  \draw [dashed,->] (2.1, -0.5) -- (0.9, 0.5) ;
  \draw [dashed,->] (0.9, 0.5) -- (2.1, 0.5) ;

  \draw (7,1)-- (10,1);
  \draw (7,0)-- (10,0);
  \draw (7,-1)-- (10,-1);
  \draw (7,1)--(7,-1);
  \draw (10,1)--(10,-1);
  \draw (8.5,1)-- (8.5,-1);
  \node at (7.8,1.2) {$h_{21}$};
  \node at (9.3,1.2) {$h_{22}$};
  \node at (6.7,0.5) {$b_{22}$};
  \node at (6.7,-0.5) {$b_{21}$};
  \node at (8.5,-1.3) {User~$2$};
  \draw [dashed,->] (9.1, -0.5) -- (7.9, 0.5) ;
  \draw [dashed,->] (7.9, 0.5) -- (9.1, 0.5) ;
  \draw [dashed,->] (7.9, -0.5) -- (9.1, -0.5) ;
    \node at (9.2, -0.5) {2};
  \node at (7.8, -0.5) {1};
  \node at (9.2, 0.5) {4};
  \node at (7.8, 0.5) {3};
  \end{tikzpicture}
\end{center}
\caption{Joint CDFs of arrivals and fading \label{fig:lex:path}}
\end{figure}
The state-pairs $(b,h)$ for each distribution can be lexicographically ordered, see
the directed paths shown in Figure~\ref{fig:lex:path}. 
Using \eqref{eq:def:level}, we can identify 
$$
(\alpha_{02},\alpha_{11}, \alpha_{12}, \alpha_{21},\alpha_{22}) 
	=\left(0,\frac 1{12}, \frac 1{6}, \frac 1{3}, \frac 12 \right)\,\, \textrm{ and}\,
(\beta_{02},\beta_{11},\beta_{12},\beta_{21},\beta_{22})
	= \left(0,\frac 18, \frac 3{16}, \frac 9{16}, \frac 3{4}\right). 
$$
These values are marked  in Figure~\ref{fig:cum:path}, where a 
dummy value $d_0 = \beta_{22} - \alpha_{22}$ was added at the base of the first 
vector to equalize the heights.

\begin{figure}[htbp]
\begin{center}
\begin{tikzpicture}[line width=1.5pt]
\draw[thin] (-0.5,0) --++(6.5,0);

\draw (5,0) rectangle ++(0.5,1);
\draw (5,1) rectangle ++(0.5,0.5);
\draw (5,1.5) rectangle ++(0.5,3);
\draw (5,4.5) rectangle ++(0.5,1.5);
\draw [thin,decorate,decoration={brace, mirror, raise=0pt}] (5.85,0) -- node[right]{$\frac 1{8}$} ++(0,1);
\draw [thin,decorate,decoration={brace, mirror, raise=0pt}] (5.85,1) -- node[right]{$\frac 1{16}$} ++(0,0.5);
\draw [thin,decorate,decoration={brace, mirror, raise=0pt}] (5.85,24/16) -- node[right]{$\frac 3{8}$} ++(0,3);
\draw [thin,decorate,decoration={brace, mirror, raise=0pt}] (5.855,72/16) -- node[right]{$\frac 3{16}$} ++(0,1.5);

\foreach \i in {0,1,1.5,4.5,6} {\draw[dashed, thin] (5,\i) --++(-3.5,0);}

\draw[dotted, thin] (5,0) --++(2.5,0) node[right]{$\beta_{02}$};
\draw[dashed, thin] (5,1) --++(2.5,0) node[right]{$\beta_{11}$};
\draw[dashed, thin] (5,1.5) --++(2.5,0) node[right]{$\beta_{12}$};
\draw[dashed, thin] (5,4.5) --++(2.5,0) node[right]{$\beta_{21}$};
\draw[dashed, thin] (5,6) --++(2.5,0) node[right]{$\beta_{22}$};

\pgftransformxshift{-2.0cm};
\draw[<->,>=stealth'] (3.25,0) -- node[left]{$d_0 = \frac 14$} ++(0,2);
\pgftransformyshift{2.0cm};
\draw (3,0) rectangle ++(0.5,8/12);
\draw (3,8/12) rectangle ++(0.5,8/12);
\draw (3,16/12) rectangle ++(0.5,16/12);
\draw (3,32/12)  rectangle ++(0.5,16/12);

\draw [thin,decorate,decoration={brace,raise=0pt}] (2.65,0) -- node[left]{$\frac 1{12}$} ++(0,8/12) ;
\draw [thin,decorate,decoration={brace,raise=0pt}] (2.65,8/12) -- node[left]{$\frac 1{12}$} ++(0,8/12) ;
\draw [thin,decorate,decoration={brace,raise=0pt}] (2.65,16/12) -- node[left]{$\frac 1{6}$} ++(0,8/6) ;
\draw [thin,decorate,decoration={brace,raise=0pt}] (2.65,32/12) -- node[left]{$\frac 1{6}$} ++(0,8/6) ;
\foreach \i in {0, 8/12,16/12,32/12,4} {\draw[dashed, thin] (2,\i) --++(5,0);}

\draw[dashed, thin] (2.5,0) --++(-1.5,0) node[left]{$d_0+\alpha_{02}$};
\draw[dashed, thin] (2.5,8/12) --++(-1.5,0) node[left]{$d_0+\alpha_{11}$};
\draw[dashed, thin] (2.5,16/12) --++(-1.5,0) node[left]{$d_0+\alpha_{12}$};
\draw[dashed, thin] (2.5,32/12) --++(-1.5,0) node[left]{$d_0+\alpha_{21}$};
\draw[dashed, thin] (2.5,4) --++(-1.5,0) node[left]{$d_0+\alpha_{22}$};

\end{tikzpicture}
\caption{Pseudo CDF-pair \label{fig:cum:path}}
\end{center}
\end{figure}
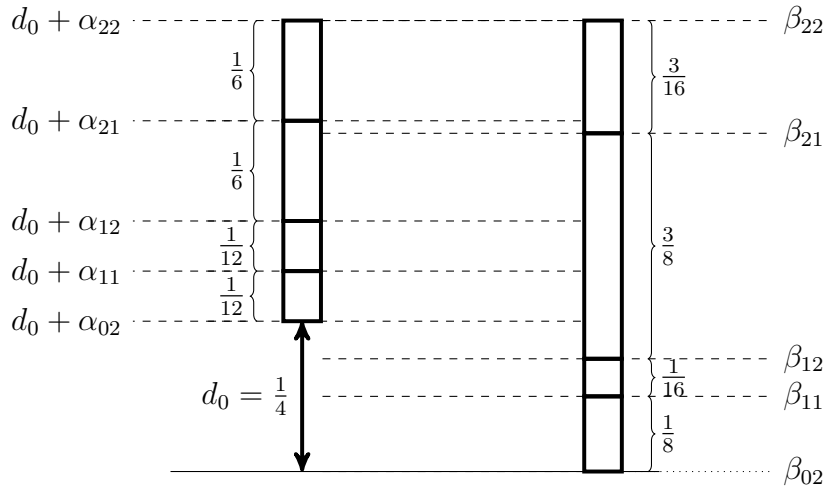

Observe that the cumulative values (labeled as $\beta_{ij}$ and $d_0+\alpha_{ij}$) shown 
in Figure~\ref{fig:cum:path} do not correspond to actual CDFs, 
we call them a pseudo CDF-pair. Notice
the dashed levels marked by horizontal lines, these values play an important role in our iterative power allocation.
The key idea which we take  forward from this example is to allocate power iteratively to each pair 
connected by a horizontal dashed level.
\end{example}

Let us  generalize this example, and lexicographically enumerate the tuples $(B_i,H_i)$ 
to construct a pseudo-CDF pair as in Figure~\ref{fig:cum:path}.
Without loss of generality, assume  $\beta_{\cBb\cHb} \geq \alpha_{\cBa\cHa}$. 
%
Using \eqref{eq:def:level}, define two maps $\chi_1$ and $\chi_2$ as follows. 
\begin{align*}
\chi_1(0,0)&=\beta_{\cBb\cHb} - \alpha_{\cBa\cHa} \\
\chi_1(B_1=b_{1j},H_1=h_{1k})&= \chi_1(0,0) + \alpha_{jk},~ 1 \leq j \leq \cBa,~ 1 \leq k \leq \cHa \\
\chi_2(B_2=b_{2j},H_2=h_{2k})&=\beta_{jk} ,~ 1 \leq j \leq \cBb,~ 1 \leq k \leq \cHb.
\end{align*}
Let $Range(\chi_i)$ denote the range of the map $\chi_i$, and take $\Gamma := Range(\chi_1) \bigcup Range(\chi_2)$,
with the elements indexed in the ascending order.  To clarify, in  Figure~\ref{fig:cum:path},
the set $\Gamma:=\{\gamma_0, \cdots, \gamma_{|\Gamma|-1}\}$  is  simply the ordered collection of the 
dashed horizontal levels shown there.
Let us also define the inverse map of $\chi_i, i=1,2$ by
\begin{align} \label{eq:inv:gamma}
\bigl( b_i(\gamma_l),h_i(\gamma_l) \bigr) = \max\{ (b_{ij},h_{ik}) : \chi_i (b_{ij},h_{ik}) \leq \gamma_l\},
\end{align}
where $\gamma_l \in \Gamma$, and the maximum is  in the lexicographical order. 
We now present an optimal power allocation scheme.
Like in Section~\ref{sec:fix:fade}, the iterative scheme proceeds in the increasing order of $\gamma_l$,
and power will be allocated at each step to the inverse of $\gamma_l \in \Gamma$, for a hitherto
unallocated pair of rate and fading-value at a user.

In the following theorem,  $P_i(b_i(\gamma_l),h_i(\gamma_l))$ is denoted as $P_i(l)$ for brevity. 
Denote the smallest index in $\{0,\cdots, |\Gamma|-1\}$ such that $\gamma_l$ corresponds to a positive rate for 
at least one of the users as $l^*$. Clearly $P_i(l)=0$ if $l < l^*$, as there is no need
for any allocation.

\begin{theorem}\label{thm:optpwr}
The power allocation functions $P_1(.)$ and $P_2(.)$ given by
\begin{align}
h_1^2(\gamma_{l-1})P_1(l-1)+h_2^2\left(\gamma_{l}\right)P_2(l)&=2^{2\left(b_1(\gamma_{l-1})+b_2(\gamma_{l})\right)}-1 \label{eq:vf:pow:1}\\
h_1^2(\gamma_{l-1})P_1(l-1)+h_2^2\left(\gamma_{l-1}\right)P_2(l-1)&=2^{2\left(b_1(\gamma_{l-1})+b_2(\gamma_{l-1})\right)}-1 \label{eq:eqlitythm} 
 \end{align}
for $l^* < l \leq |\Gamma|-1$, with the initial power allocation 
satisfying
\begin{align}
 h_1^2(\gamma_{l^*})P_1(l^*) +h_2^2(\gamma_{l^*})P_2(l^*) &=2^{2\left(b_1(\gamma_{l^*}) + b_2(\gamma_{l^*})\right)}-1 \\
h_i^2(\gamma_{l^*})P_i & \geq 2^{2b_i(\gamma_{l^*})}-1, \, i =1,2, \label{eq:vf:pow:2}
\end{align}
achieve
\begin{align}
\eE P_1(B_1,H_1) + \eE P_2(B_2,H_2) = P_{avg}^{min}(1).
\end{align}
\end{theorem}
\begin{IEEEproof}
The proof can be found in Appendix~\ref{sec:optpwr}.
\end{IEEEproof}
It now remains to be shown that the power allocation scheme in Theorem~\ref{thm:optpwr} is outage free. 
\begin{lemma} \label{lem:outage:2}
The power allocation given in \eqref{eq:vf:pow:1} -- \eqref{eq:vf:pow:2} is an outage free scheme
over a distributed MAC with bursty arrivals.
\end{lemma}
\begin{IEEEproof}
The proof is given in Appendix~\ref{sec:outage:2}.
\end{IEEEproof}
We have thus shown an optimal scheme which achieves $P_{avg}^{min}(1)$, and is outage free. Before
embarking on a simulation study, some comments are in order.
It should be noted that the channel values  are not ordered monotonically 
while constructing the pseudo-CDF pair (see Figures~\ref{fig:lex:path}-\ref{fig:cum:path}), 
it is enough to take the required transmission rates at each user in the increasing order
 while the powers are iteratively assigned. In particular, the fading values and 
their probabilities play a role in the construction of the pseudo-CDF pair. 
\begin{remark} \label{rem:var:fade}
Suppose that after evaluating the pseudo-CDF pair, we replace every fading value by unity. 
The power allocation in Theorem~\ref{thm:optpwr} will now specify the 
required received power for each  transmission-rate chosen by a user. 
Clearly, the transmit powers at the CeNs of the original MAC can be found by appropriate scalings. 
\end{remark}
Notice that for each $\gamma_l \in \Gamma$,  \eqref{eq:inv:gamma} defines a pair of values 
at user~$i \in \{1 \leq i \leq L\}$,
let $I_l$ denote the ordered collection of these $L$ pairs. 

\begin{remark} 
The knowledge of the set $\{I_l, 0 \leq l \leq |\Gamma|-1\}$ at each user is sufficient
to specify the  complete power-allocation scheme. Thus, even the knowledge of the statistics 
is redundant while designing the communication scheme, once the  users have access to $\{I_l,
0\leq l \leq |\Gamma|-1\}$.
\end{remark}

An astute reader might have observed that our approach in this section differed slightly
from the exposition in Section~\ref{sec:fix:fade}. 
While the marginal CDFs of the arrivals were used in
the power allocations of Section~\ref{sec:fix:fade}, we employed pseudo-CDFs here. 
The latter approach saved us from an explosion of notations in the presence of
dynamic fading. 
We now detail how Lemma~\ref{lem:outage:2}
will imply Lemma~\ref{lem:achieve:1}, this will also explain the equivalence of the two 
approaches.

Suppose we have a fixed fading MAC with the respective 
fading power gains $\alpha_1$ and $\alpha_2$ with $\alpha_1 \geq \alpha_2$. Let $\phi_1$
and $\phi_2$ denote the respective marginal CDFs of the rate arrivals.
If $P_{\mathrm a}^{\mathrm m}(\alpha_1, \alpha_2)$ denotes the minimum average sum-power for this MAC
with bursty arrivals under a unit slot delay constraint, then
$$
P_{\mathrm a}^{\mathrm m}(\alpha_1, \alpha_2) = \alpha_2 P_{\mathrm a}^{\mathrm m}\left(\frac{\alpha_1}{\alpha_2},1\right).
$$
Thus, we can equivalently find the minimal average power for a two user MAC with  
fading power gains $(\frac{\alpha_1}{\alpha_2},1)$. For the latter channel, suppose we arrange the values of 
$\phi_1$ and $\phi_2$ as in Figure~\ref{fig:cum:path}, and generate the ordered set $\Gamma$.
Clearly, $d_0 = 1 - \frac{\alpha_2}{\alpha_1}$ and  $\beta_{|\mathcal B_2||\mathcal H_2|} =1$, 
i.e. the height of the graph is unity.
 Furthermore, using \eqref{eq:inv:gamma}
$$
\{(b_1(\gamma_l),b_2(\gamma_l)), 0 \leq l \leq |\Gamma| -1\} =  
	\{\left(\phi_1^{-1}\bigl(\frac{\alpha_1 (x - d_0)}{\alpha_2}\bigr), \phi_2^{-1}(x)\right), 0 \leq x \leq 1\}.
$$
Observe that the RHS is exactly the set of rate-pairs for which Lemma~\ref{lem:achieve:1} 
allocated the minimum required transmit power.
Thus, for each $0\leq l \leq |\Gamma|-1$, the power allocation for the pair $(b_1(\gamma_l),b_2(\gamma_l))$
is identical in  Lemma~\ref{lem:achieve:1} as well as Lemma~\ref{lem:outage:2}. Therefore, the allocation
in Lemma~\ref{lem:achieve:1} is a special case of Lemma~\ref{lem:outage:2}.

\subsection{Simulation Study} \label{sec:sim:2}

Let us now compare the performance of the proposed schemes with TDMA
as well as centralized schemes. A generalized
TDMA scheme (G-TDM) is used in the simulations below for comparisons, 
where the fraction of the time given to a user is optimized to get the maximum time-shared sum-rate. 
The optimal centralized scheme is as follows.

\noindent \emph{Centralized Scheme}:
In a centralized scheme, each user knows the global CSI as well as the
rate-requirements at all terminals.
While \eqref{eq:pow:contra} still needs to be satisfied for each rate-vector, one
can achieve equality in that equation, thus reducing the required average transmit-power 
in comparison with a decentralized system.
 %
 %
%
With the channel coefficients $(\sqrt{\alpha_1}, \sqrt{\alpha_2})$, 
minimum transmit sum-power to support the rate-tuple
$(b_1,b_2)$ in a slot can be evaluated as 
%
\begin{align} \label{eq:cent:opt}
 \min P_1+P_2 \text{ subject to: } {\sum_{i \in J} \alpha_i}P_i \geq {2^{2(\sum_{i \in J}b_i)}-1}, 
	\forall J \subseteq \{1,2\}.
  \end{align}
The feasible power-pairs which can support the rate-pair $(b_1,b_2)$ is a
contra-pentagon, similar to that shown in Figure~\ref{fig:contra}. Clearly \eqref{eq:cent:opt}
can be solved by operating at the corner-points of the contra-pentagon. In particular,
the optimal operating point is always chosen from the line 
 $\alpha_1P_1+\alpha_2 P_2={2^{2(b_1+b_2)}-1}$. If $\alpha_1 < \alpha_2$ we can take
$\alpha_1 P_1 = 2^{2b_1}-1$, otherwise we take $\alpha_2 P_2 = 2^{2b_2}- 1$.
Notice that if $\alpha_1=\alpha_2$, one can operate anywhere on the dominant face.

%
%

In the first simulation below, the effect of variations in the fading
statistics on the total power consumption is studied. 
%
Let $H_2$ be uniformly distributed
in $\{1,\cdots,5\}$, and $H_1$ be uniformly distributed in 
$\{\gamma_a, 2 \gamma_a, 3 \gamma_a, 4 \gamma_a, 5 \gamma_a\}$, where $\gamma_a$ is a positive
parameter capturing the asymmetry in the links for the two users. 
%
Assume that the arrivals for user~$i\in\{1,2\}$
are chosen with probability
\begin{align} \label{eq:arrive:law:1}
 Pr(B_i=k-1)=\frac{p_i{(1-p_i)^{(k-1)}}}{[1-(1-p_i)^{5}]}, \, 1 \leq k \leq 5.
\end{align}
Notice that $B_i+1$ is a truncated Geometric distribution.
The parameter $p_2$ is taken to be $0.25$ for all the numerical computations below.
%

\begin{figure}[htbp]
\begin{center}
\includegraphics[scale=1.0]{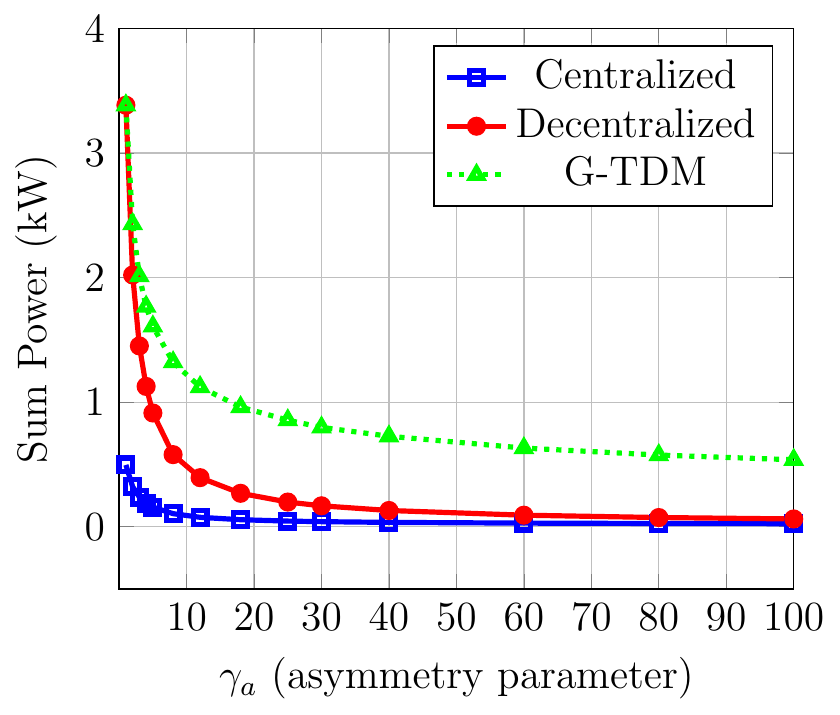}
   \caption{Decentralized schemes vs TDMA and centralized schemes, $D_{max}=1$~\label{fig:var:fade}}
\end{center}
\end{figure}

Figure~\ref{fig:var:fade} compares the average sum-power expenditure 
when the link asymmetry parameter 
$\gamma_a$ is varied from $1$ to $100$, while keeping $p_1=p_2=0.25$. Clearly, 
when the statistical laws are identical at both the users, the decentralized system and 
G-TDM give similar performance, whereas
there is a lot to be gained by centralized operations. However, as the fading laws
become more asymmetric, the optimal decentralized schemes perform superior to G-TDM. 

Let us now study the effect of variability in arrival distributions as well.   
Let $H_2$ be uniform in $\{1,2,3,4,5\}$, and $H_1$ be independently and uniformly
taken from $\{\gamma_a, 2\gamma_a,\cdots, 5\gamma_a\}$.
Let us  fix the parameter $p_2$  in \eqref{eq:arrive:law:1} at $0.25$, and vary
$p_1$ in an appropriate range.
%

\begin{figure}[htbp]
\begin{center}
\includegraphics[scale=1.0]{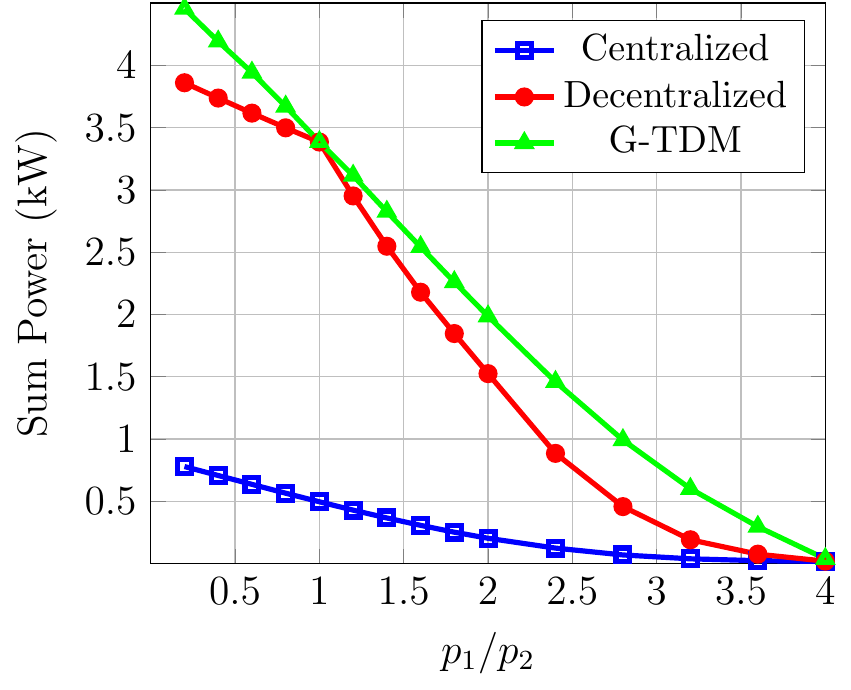}
   \caption{Sum power versus $p_1/p_2$, with the probability parameter $p_2=0.25$, asymmetry parameter $\gamma_a=1$.\label{fig:var:arrive:1}}
\end{center}
\end{figure}

\begin{figure}[htbp]
\begin{center}
\includegraphics[scale=1.0]{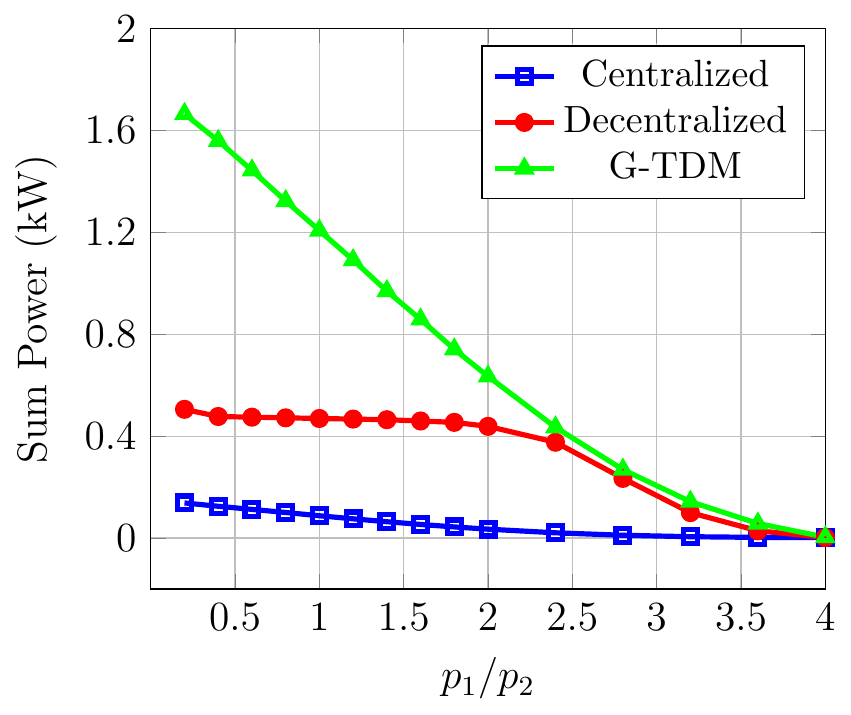}
   \caption{Sum power versus $p_1/p_2$, with the probability parameter $p_2=0.25$, asymmetry parameter $\gamma_a=10$.\label{fig:var:arrive:2}}
\end{center}
\end{figure}

Figures~\ref{fig:var:arrive:1} and \ref{fig:var:arrive:2} plot the average sum-power as a function of 
the ratio $p_1/p_2$ 
for $\gamma_a=1$ and $\gamma_a=10$ respectively. Note that
for $\gamma_a=1$ and $p_1=p_2$, the two users are statistically identical and hence the decentralized
scheme has performance similar to G-TDM.
As the ratio $p_1/p_2$ increases, the probability of lower sized packets  at 
user~$1$ increases, hence the required average sum-power diminishes for all the schemes. However, it is
evident that the proposed scheme outperforms G-TDM.
Similarly, for $\gamma_a=10$, the decentralized scheme is almost identical to TDMA when 
$p_1/p_2 \approx 2.8$,  but has  superior performance in other ranges.

\section{Distributed Scheduling Under a  General Max-delay Constraint} \label{sec:del:dmax}
So far we have considered a distributed  MAC with bursty arrivals under a unit slot 
delay constraint. A unit-slot delay is a very stringent
requirement, relaxed QoS guarantees are more applicable. Let us now
consider the widely employed max-delay constraint, i.e. each packet should be delivered before 
$D_{max}$ slots, where $D_{max} \geq 1$ is some specified integer~\cite{Khoj04}. 
While we can  also allow a separate max-delay constraint for each queue, this will only add 
notational burden. Since our primary motivation is to analyze the relaxation of delay
requirements, we will consider a MAC with fixed fading coefficients and bursty arrivals 
in this section.

It was already shown in Section~\ref{sec:model} that  the operations of 
BiS and CeN can be decoupled at each transmitter (see Figure~\ref{fig:tandem}). 
More specifically, the CeN $P_i, i=1,2$ operates under
a unit delay constraint on the scheduled bits from its corresponding BiS~$S_i$. 
Furthermore, each CeN encounters a stationary ergodic arrival process, as opposed to the IID
inputs considered in the previous section. As observed in Remark~\ref{rem:erg:ext}, this can 
be readily handled by the power allocations in Lemma~\ref{lem:achieve:1}, by using the 
stationary marginal CDFs there.  Furthermore, Remark~\ref{rem:rate:pow} enables us to 
construct a suitable rate-power characteristics $P_l(b), 0\leq b \leq |\mathcal B_l|$ for user~$l\in\{1,2\}$.

\begin{figure}[htbp]
\begin{center}
\begin{tikzpicture}[scale=0.95]
\begin{axis}[xlabel={$b$, bits/s/Hz}, ylabel={$P_1(b)$, Watts},grid]
\addplot[line width=2.0pt, mark=*, blue] coordinates{(0,0) (1,19.2) (2,96) (3,403) (4,1632)};
\end{axis}
\end{tikzpicture}
\vspace*{-0.5cm}
\end{center}
\caption{Rate-Power Characteristic at CeN~$P_1$\label{fig:pow:conv}}
\vspace*{-0.25cm}
\end{figure}
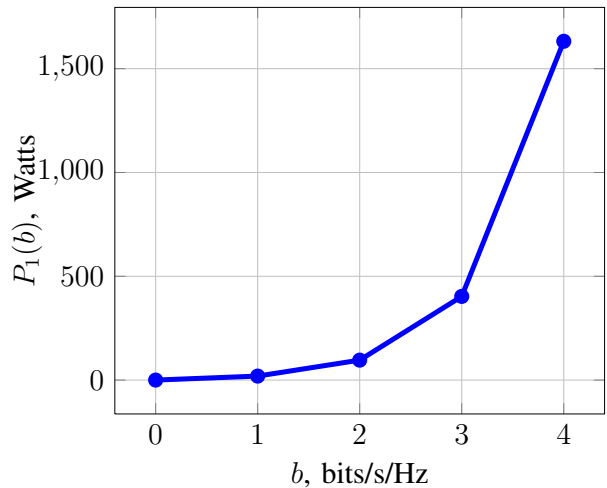

Figure~\ref{fig:pow:conv} illustrates the rate-power curve for one of the schedulers specified
in Figure~\ref{fig:mat1}, where we have
taken $\alpha_1=10, \alpha_2=1$ and uniform arrivals in $\{1,2,3\}$. 
The power allocation $P_1(b_1)$ for 
rates $b_1 \in [0,4]$ is shown, where $\{B_1=4\}$ is an additional dummy state. 

%
%

The following local relationship is immediate in lieu of Remark~\ref{rem:rate:pow}.
\begin{claim} \label{claim:opt:su}
For an optimal outage free communication scheme $(\bar\bS,\bar\bP)$ at the transmitters, 
the scheduler $S_i$ at BiS~$i$ is an optimal single-user scheduler for  
the  power allocation function $P_i(\cdot)$.
\end{claim}
\begin{IEEEproof}
Assume on the contrary that some $(S_i,P_i)$ does not meet the asserted property. By keeping
all other schedulers and power allocations  the same, we can decrease the 
average sum-power by choosing an optimal $S_i$  for the given $P_i$.
\end{IEEEproof}
%
Let $S_{\textrm{su},i}(P_i)$ denote the optimal single user stationary scheduling
policy when the rate-power characteristic at terminal~$i$ is given by the function $P_i(\cdot)$. 
Optimal single user scheduling is  a reasonably well understood topic \cite{Rajan04},  
\cite{Khoj04}, typically solved by dynamic programming, see \cite{Bertsekas2001} for a 
detailed exposition and relevant examples. Using the 
optimal single schedulers $S_{\textrm{su},i}(\cdot)$, we now present an iterative algorithm to evaluate the
optimal average sum-power required to successfully transport the arriving data in a 
distributed fashion.

\subsection{Optimal Scheduling Algorithm}

\begin{center}
\begin{tikzpicture}
\node at (0,0) [text width = 16.0cm, draw, rectangle,rounded corners=4pt]{
\noindent \textbf{Algorithm~IterOpt}

\noindent\begin{enumerate}
  \item[1:]The initial power policy $\bar\bP$ is taken as the optimal unit slot delay allocation.
  \item[2:]For $\bar\bP$, find the
	 optimal single user stationary schedulers $S_{\textrm{su},i}(P_i), i=1,2$.
  \item[3:] Perform optimal unit slot delay power allocation 
 for the new set of marginal rate distributions at the BiSs.
  \item[4:] Go back to Step~$2$ using the power allocations from the last step.
 \end{enumerate}
};
\end{tikzpicture}
\end{center}
Algorithm~IterOpt is terminated when the required average sum-power becomes invariant. 
Notice that we are performing an alternate minimization or Gauss-Siedel minimization
on a convex (not strictly) utility~\cite{Beck2014}. Interestingly, 
in spite of not having strict convexity, the algorithm is guaranteed to converge  to 
the optimal value, when optimized over the set of  schedulers $\bar{\mathcal S}$ 
meeting the maximal delay constraint. 
Let $P^{*}_{HALT}$ be the terminal  average sum-power given by Algorithm~IterOpt.
\begin{prop} \label{prop:algo:conv}
Algorithm~IterOpt terminates by achieving the optimal average sum-power, i.e. we have
$P^{*}_{HALT} = P_{avg}^{min}(D_{max})$.
\end{prop}
\begin{IEEEproof}
The proof is given in Appendix~\ref{sec:prop:proof}.
\end{IEEEproof}
Step~2 of the algorithm required the availability of 
optimal single user schedulers $S_{\textrm{su},i}(\cdot)$ for each of the given 
convex rate-power characteristics.
This involves solving a DP similar to~\cite{Rajan04} at each BiS, where 
computational approaches seem necessary.

\subsection{Single User Scheduling}\label{sec:mdp}
 Recall that for a given power function $P_i(\cdot)$ and buffer state 
 $\zeta_i[j]$ (see Definition~\ref{def:state:vec}), 
 the BiS~$S_i$
 decides an optimal action by choosing an appropriate transmission rate~$r$ for slot~$j$. 
 In Algorithm~IterOpt, we indeed  assumed the availability of an optimal single user
 scheduler.
 The optimal scheduling policy is identified typically by dynamic programming 
 approaches~\cite{Bertsekas2001}. 
 While closed form solutions are not always available, 
 a computational approach known   as value iteration algorithm (VIA) can numerically determine the
 optimal schedules, by solving the Bellman equation for the corresponding discounted
 cost problem given by
 \begin{equation} \label{eq1}
    V_{j+1}(s)=\min_a \{P(a) + \sum_{s^\prime}\gamma Pr(s^\prime | s,a)V_j(s^\prime)\} .
 \end{equation}
  Here, $j$ denotes the iteration number, $s$ is the $D_{max}$ dimensional vector of the current buffer state,   
and  $P(a)$ is the power required for the action (transmission-rate) $a$.
The function $Pr(s^\prime | s,a)$ is the probability of buffer going from state $s$ to 
  state $s^\prime$ under the action  $a$,  and $\gamma$ is a discount factor, taken slightly below unity.

In the VIA,
the scheduled rate $a$ can take any value from $[s_1, \sum_{i=1}^{D_{max}} s_i]$,
in steps of $\Delta$, which is the step-size parameter. The step-size can be chosen 
appropriately to improve either the speed or accuracy.  
In particular, integer-valued schedulers can be obtained by setting $\Delta=1$. 
Note that the objective function is non-decreasing with $\Delta \in (0,1]$.
Since $P(a)$ is convex in action~$a$ (see Lemma~\ref{lem:conv:rate}), 
the VIA will converge for each $\Delta$, 
specifying the optimal scheduler for the power allocation function at each user.
We now illustrate Algorithm~IterOpt by an example.

\begin{example}
Let us take $D_{max}=2, \alpha_1=10, \alpha_2=1$, and
assume both the arrivals to be uniform in $\mathcal A = \{1,2,3\}$.
We can start with the initial schedulers as shown in Figure~\ref{fig:mat1}, which are designed
using a TDMA based power-allocation. Using a step-size of $\Delta=1$ (integer-valued schedulers), 
Algorithm~IterOpt outputs the schedulers $S_1^{final}$ and $S_2^{final}$ 
shown in  Figure~\ref{fig:mat2}, after two iterations.
\begin{figure}[htbp]
 \[ 
S_1^{final} =\begin{blockarray}{cccc}
&  1 & 2 & 3  \\
\begin{block}{c [ccc]}
  0 & 1 & 2 & 2 \\
  1 & 2 & 2 & 2 \\
  2 & 2 & 2 & 3 \\
  3 & 3 & 3 & 3\\
  \end{block}
\end{blockarray} \qquad
S_2^{final} =\begin{blockarray}{cccc}
&  1 & 2 & 3  \\
\begin{block}{c [ccc]}
  0 & 1 & 2 & 2 \\
  1 & 2 & 2 & 2 \\
  2 & 2 & 2 & 2 \\
  3 & 3 & 3 & 3\\
  \end{block}
\end{blockarray}
\]
\caption{Schedulers $S_1$ and $S_2$ after iterations \label{fig:mat2}}
\end{figure}

\end{example}

\subsection{Simulation Study}\label{sec:sim:3}
We now demonstrate the advantages of using the proposed iterative
power minimization framework over conventional TDMA-based schemes, or the 
robust scheduling framework of \cite{Khoj04}.  The available slot is equally
shared between the users in the TDMA scheme employed for comparisons here. 
The examples
below are taken to be simple enough, yet they capture the intrinsic operational details,
and expected performance enhancements. Let us consider a two user  MAC system with 
fixed channel values of $1$ and $\sqrt{\alpha}$ respectively. We take arrivals to be uniform in
$\mathcal A = \{1,2\}$ for our experiments.
\subsubsection{Integer-valued Schedulers}
\begin{figure}[htbp]
  \centering
  \includegraphics[scale=1.0]{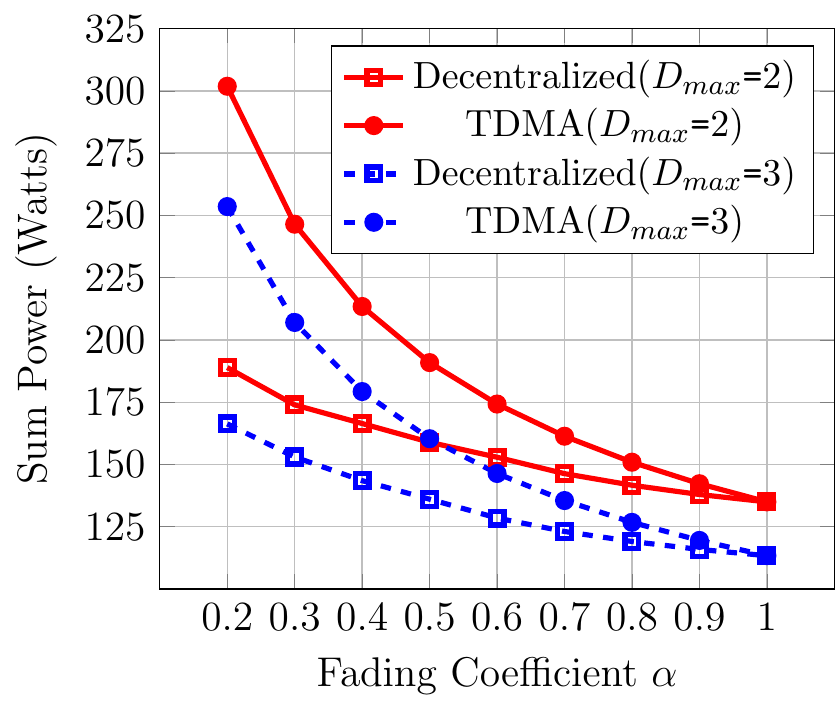}
  \caption{Integer BiS+Optimal CeN  Vs Integer BiS+TDMA,  $\alpha_1=1, \alpha_2 = \alpha$.}\label{fig:plot1}
   \end{figure}

Recall that schedulers with integer-valued rate outputs can be obtained by 
setting $\Delta=1$ in the VIA, starting from any integer scheduler. 
We  compare the performance of the scheduler obtained by our iterative 
algorithm to the one using TDMA in conjunction with the optimal single
user integer schedulers, see \cite{Rajan04} for the latter.
The average sum-power is plotted as a  function of the link parameter $\alpha$ in
Figure~\ref{fig:plot1}.  
Observe that the proposed strategy and TDMA performs equally well when $\alpha=1$, i.e.
when the conditions at both users are identical.
But when $\alpha$ moves away from $1$, the advantage of 
using the strategies proposed in this paper is evident.

\subsubsection{Robust Schedulers with Optimal Power Allocation}
\begin{figure}[htbp]
\begin{center}
\includegraphics[scale=1.0]{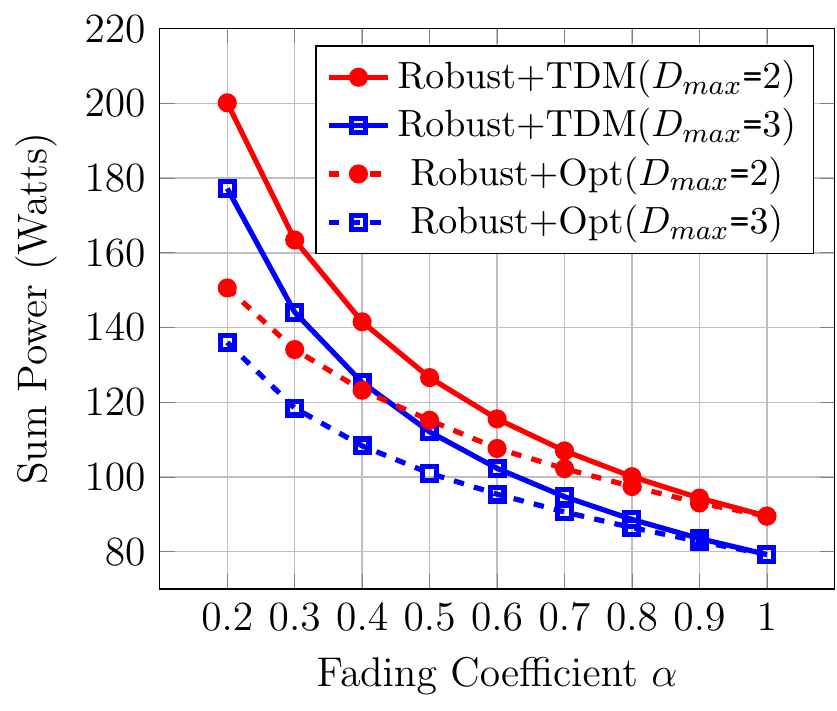}
\end{center}   

\caption{Robust Schedulers at BiS with Optimal/TDMA Power Allocation, $\alpha_1=1,\alpha_2=\alpha$. }\label{plot2}
   \end{figure}

We now show that the performance improvement with respect to TDMA is visible even in
rational (non-integer) scheduling setups. In particular, we show that even 
if one commits to the robust schedulers of \cite{Khoj04} at the BiSs, the power efficiency 
of the allocation in Lemma~\ref{lem:achieve:1} is superior to the non-integer schedulers based on TDMA.
Notice that the robust schedulers are agnostic to the arrival distribution~\cite{Khoj04}.
Figure~\ref{plot2}
compares the power expenditure when the link parameter ${\alpha}$ is varied
form $0.2$ to $1$ for $D_{max}=2$ as well as $D_{max}=3$. 
With reference to Figure~\ref{plot2}, a robust time-varying scheduler in conjunction with power 
allocations
of Lemma~\ref{lem:achieve:1} can be a reasonable choice for  distributed scheduling 
in a MAC with bursty arrivals. 
\subsubsection{Robust Scheduling Vs Optimal Scheduling}
\begin{figure}[htbp]
\begin{center}
\includegraphics[scale=1.0]{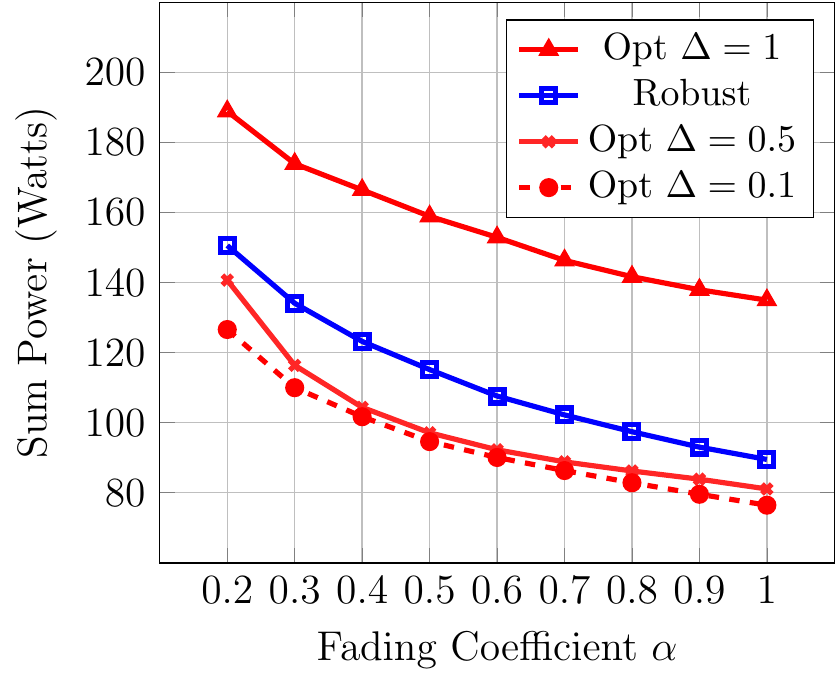}
\end{center}
   \caption{Performance of schedulers with variable step sizes for $D_{max}=2$}\label{plot3}
   \end{figure}

Let us now design optimal (real-valued) schedulers using the VIA at 
different step sizes, say $\Delta=0.5$ and  $\Delta=0.1$, 
as explained in Section~\ref{sec:mdp}. For $D_{max}=2$, 
Figure~\ref{plot3} shows the average sum power of real-valued schedulers at these step sizes, 
used in conjunction with the optimal power laws of Lemma~\ref{lem:achieve:1}. 
It can be seen that with a step size $0.5$ and less, the proposed scheduler outperforms the robust 
scheduling framework. Thus, the knowledge of arrival statistics can be put to good use
by appropriately factoring these in the dynamic program. Notice also that
the performance of a real-valued scheduler may further improve with a reduction in 
the VIA step size.

\subsection{Complexity of Algorithm~IterOpt}
Notice that Algorithm~IterOpt needs to be run only once at the start, before the transmissions
begin. Using
the arrival statistics  from all the users, the algorithm specifies 
a BiS and a CeN at each transmitter. Since the iterative procedure
is to be done only once for a given set of statistics, some level
of computational complexity is acceptable, and can be amortized over time.
It is reasonable to assume that a single entity computes the communication
scheme before any transmission starts, and supplies the relevant 
rate and power allocation functions
to all the terminals. On the other hand, it is also possible for
each terminal to separately run Algorithm~IterOpt using the available
statistics. In the latter case, the complexity mentioned below needs to be
scaled by the number of users.

The most computationally intensive part of  the Algorithm is Step~$2$.
This involves solving a dynamic program (DP).  While
closed form solutions are not often available for DPs, approximate
solutions are obtained using value iteration or policy iteration~\cite{Bertsekas2001}. 
As pointed
out in \cite{Neely07}, solving MDPs with multiple queues usually leads to
a complexity explosion. However, in our algorithm, each terminal solves
a separate MDP, and  the queues do not interact under a given power allocation
scheme. Thus the complexity is linear in the number of users. 
Given the arrival processes,  the
number of possible buffer states $M$ and  number of possible actions $N$ at each user
are determined by the choice of the quantization level $\Delta$. The VIA used in
our simulations is of polynomial complexity
in both $M$ and $N$. Thus, very fine quantizations and/or higher values of 
$D_{max}$ can make the computations intractable. However, there are ways to speed
up the MDP computations at the expense of accuracy. 
In any case, solving the MDP or finding approximate/heuristic 
solutions thereof seems an unavoidable step in  communication schemes
minimizing the average transmit sum-power under delay constraints~\cite{Rajan04}.

Step~$3$ of Algorithm~IterOpt is also polynomial in the number of states, as it 
solves for the stationary distribution of a Markov Chain. 
For the iterative power allocation scheme, power  needs to be assigned
once for each and every rate at a terminal, thus the complexity of power allocation 
is linear in the number of scheduled rates at each terminal. Clearly, we have effectively
used  the individual arrival statistics in formulating the MDP, whereas 
the global statistics were used in specifying the power allocation.

\section{Conclusion}\label{sec:conc}
In this paper, we presented optimal multiuser communication schemes for the 
transmission of independent bursty traffic over a distributed multiple access 
channel under a max-delay constraint. An iterative algorithm was proposed to 
evaluate the minimum average sum-power. While results
are given for a two user model, generalizations to $N$ users are possible.
The unit slot delay power allocation of Section~\ref{sec:fix:fade} is the key 
to such extensions, as the rest of the results are largely user independent,
except for the computational requirements. The many user unit slot delay
power allocation for static fading, for example, can be obtained as outlined
below. 

Let $\alpha_1, \cdots, \alpha_L$ be the fading power gains in the descending
order. 
Recall that $b_i(x)$ is the inverse CDF function for the rates arriving 
at user~$i$. Suppose we scale each probability at user~$i\in\{1,\cdots, L\}$ by 
$\frac{\alpha_L}{\alpha_i}$ and place an additional probability mass of 
value $1-\frac{\alpha_L}{\alpha_i}$ at zero, to obtain a transformed CDF $\psi_i$.
Let $\Gamma:=\{\gamma_0,\cdots, \gamma_{|\Gamma|-1}\}$ be the ordered
union of the range of $\psi_i, 1\leq i \leq L$.

Denote $\hat b_i(l) := b_i\left(  (\gamma_l - 1 + \frac{\alpha_L}{\alpha_i} )\frac{\alpha_i}{\alpha_L}\right)$
and $P_i(l):= P_i(\hat b_i(l))$.
Let us now iteratively allocate powers to the rate-tuples 
$\hat b_1(l),\cdots, \hat b_L(l)$ in such a way that
$$
\sum_{i=1}^L \alpha_i P_i(l) = 2^{2\sum_{i=1}^L \hat b_i(l)} - 1.
$$
In particular, the allocation
\begin{align}
\alpha_i P_i(l) = 2^{2(\sum_{j=1}^{i-1} \hat b_j(l-1) + \sum_{j=i}^L \hat b_j({l})) } - 1 
	- \sum_{j=1}^{i} \alpha_j P_j (l-1) - \sum_{j=i+1}^L \alpha_j P_j(l),
\end{align}
will do the job, starting with an appropriate initial power allocation
on the dominant face of the corresponding contra-polymatroid. Notice that 
this allocation assigns a power to each rate at every user.


Intuitively, each BiS attempts to smoothen the traffic, in such a way
that the transmit power is kept steady across slots. In the absence
of fading, considerable smoothening can be achieved by even simple techniques
such as sending fractions of size $1/D_{max}$ of a packet for $D_{max}$ consecutive
slots. The iterative power allocation will now specify 
the optimal transmit powers. However, more care is required in presence of fading.
While Remark~\ref{rem:var:fade} helps here,  extending
the optimal schemes in Section~\ref{sec:del:dmax} to both time-varying fading as well
as arrivals, under a general max-delay constraint, appears difficult. 
In presence of fading, even a single user optimal BiS becomes 
more complicated to solve. This difficulty can be sidestepped by taking recourse
to efficient, but suboptimal, scheduling heuristics at the BiS.  
We demonstrate the performance of adapting  a 
heuristic policy for the point to point channel  from \cite{ChMiNe07}, to our MAC model.

\begin{figure}[htbp]
\begin{center}
\includegraphics[scale=1]{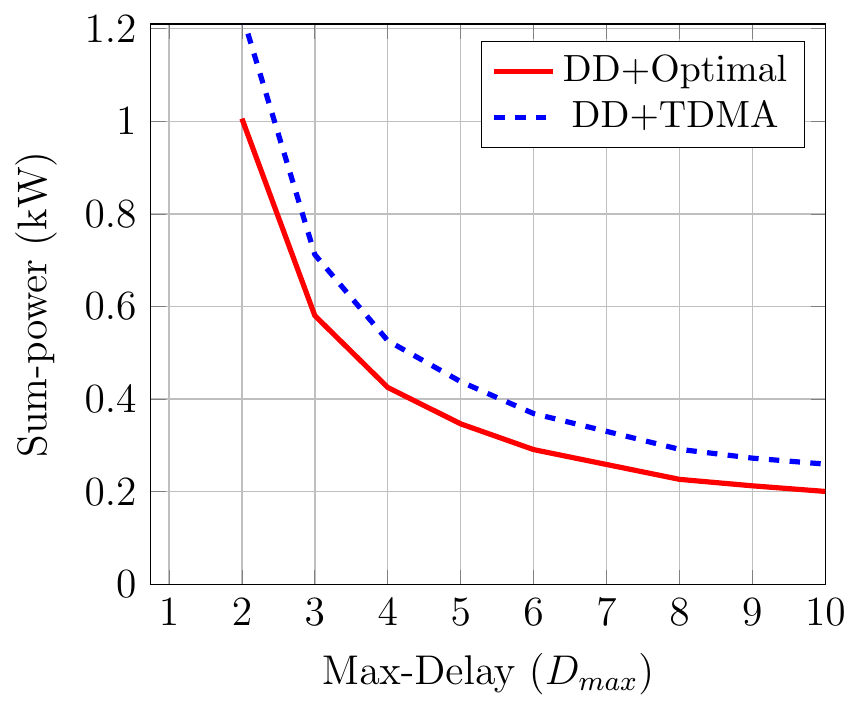}
\end{center}
\caption{Dynamic Fading and Arrivals~\label{fig:dmax:fade}}
\end{figure}

Assume a distributed model where each transmitter is aware only of its own arrivals and
time varying fading parameters. Let us employ the
\textit{Derivative Directed} (DD) online adaptive scheduler proposed by \cite{ChMiNe07} at
each BiS. This mimics a
water-filling scheme by attempting to maintain the derivative of the power allocation
at terminal~$i$. 
For a given power allocation function $P_i(r,h)$ at user~$i$ in slot~$j$, 
we compute an estimate $D_i[j]$ of  the derivative of the power allocation with
respect to the rate~$r$, where $h$ is the  fading power gain. 
A rate value $r_i[j]$ is now  chosen such that 
$$
P_i^{\prime}(r,h_i[j]) = D_i[j].
$$
Furthermore, for the buffer state vector $\zeta_i[j]$
at BiS~$i$, the transmission rate $B_i[j]$ in slot~$j$ is taken as
\begin{align}
B_i[j] = \min \Bigg\{ \max\bigg\{r_i[j] , \max_{1\leq d \leq D_{max}} \frac 1d {\sum_{k=1}^d \zeta_i[k]} \bigg\}, 
	\sum_{d=1}^{D_{max}} \zeta_i[d]  \Bigg\}.
\end{align}
The  derivative estimate is updated  in each slot using
\begin{align}\label{update_deri}
 D_i[j]= \beta D_i^{\prime}[j-1] + (1-\beta) P_i^{\prime} \Big(B_i[j],h_i[j] \Big),  ~~~~~~\text{where }0 < \beta \leq1.
 \end{align}
This scheduling scheme meets the maximum delay constraint.
Also, the optimal power allocation for $D_{max}=1$ from Section~\ref{sec:var:fade} 
gives a convenient starting point. Figure~\ref{fig:dmax:fade} 
below compares the performance of DD online scheduler under equal fraction TDMA and the optimal power allocations,
 for an example where the arrivals are uniform in  $\mathcal A = \{0,1,\cdots,4\}$, and the fading
coefficients $H_1$ and $H_2$ are chosen uniform in $\{3,4\}$ and $\{1,2\}$ respectively.

While we chose a single delay constraint for all the users, the results are expected to 
hold under different max-delay constraints at the transmitters.
Identifying
the optimal communication schemes for an average packet-delay constraint is 
an interesting future-work. Throughput maximization 
under energy harvesting nodes in a MAC~\cite{KP16} appears to have some dual relations with
the average power minimization problem here.
Exploring this duality is another future work. Lastly, we have put the knowledge
of the arrival statistics to good use in solving the decentralized MAC problem. 
In principle, one can start with any outage free communication scheme and possibly
learn some of statistical parameters from the available resources, 
using  online learning algorithms~\cite{Bertsekas2001}. 
This can then be used to progressively update the schedulers.
This scheme, in fact, builds on the proposed solutions here, and will be
explored further in future. 

While we have stated the results for minimum average sum-power,
the CDF transformation technique in \cite{SDP15jrnl} can be applied here to 
evaluate the minimum weighted average sum-power also. 


\begin{appendices}
\section{Proof of Lemma~\ref{lem:reform}} \label{sec:app:reform}
It is known that for a MDP formulation with 
bounded costs and finite state-space, there exists a deterministic 
stationary Markov policy which is average cost optimal~\cite{Sennott99}, \cite{LermaLasserre96}. 
Since we assume bounded arrivals and a maximal delay constraint in our
model, the queue-states have bounded entries as well. 
The essential idea of the proof now is to employ a quantization of the state-space.
\begin{IEEEproof}
 We assumed the transmit power
at each terminal to be continuous in the data-rate requirement.
Thus for any required transmission rate $r$, adding a
\emph{dummy} rate of $\epsilon>0$ will cause the required transmit power at that terminal to increase by
at most  $\delta(\epsilon)$,
with $\delta(\epsilon) \rightarrow 0$ as $\epsilon \rightarrow 0$. Note that the utility in
\eqref{eq:emp:power}  is normalized with respect to the number of slots $M$. Thus, adding dummy rates
of size at most $\epsilon$ to each state vector will increase the empirical
average power requirement by an amount less than  $L \delta(\epsilon)$, which is negligible for
small enough $\epsilon$. 

Observe that for  $\epsilon>0$, the state-space is discrete with bounded entries.
In this case, a deterministic stationary policy solves the
average cost MDP formulation~\cite{Sennott99}. Thus we can limit our search to deterministic 
stationary policies, where
a terminal's scheduling decision is entirely
determined by its state-vector, independent of the time of its occurrence. Notice that
the discretization makes the state-space and action space finite, implying the
existence of the limit in \eqref{eq:det:power}.

\end{IEEEproof}

\section{Proof of Theorem~\ref{thm:optpwr}} \label{sec:optpwr}
\begin{IEEEproof}
 Let us first find a lower bound to the average power. Take $P_1(0,0)=0$, and $\gamma_{-1}=0$.
\begin{align}
&  E_{\phi_1,\psi_1}\left(P_1(B_1,H_1)\right)+E_{\phi_2,\psi_2}\left(P_2(B_2,H_2)\right) \notag \\
&=\sum_{j=1}^{\cBa}\sum_k P_1(b_{1j},h_{1k})p_{1j}q_{1k}+\sum_{j=1}^{\cBb}\sum_k P_2(b_{2j},h_{2k})p_{2j}q_{2k} \notag \\
&=\sum_{j=1}^{\cBa}\sum_k h_{1k}^2P_1(b_{1j},h_{1k})\frac{p_{1j}q_{1k}}{h_{1k}^2} +\sum_{j=1}^{\cBb} \sum_k h_{2k}^2P_2(b_{2j},h_{2k})\frac{p_{2j}q_{2k}}{h_{2k}^2} \notag \\
&=\sum_{j=1}^{\cBa}\sum_k h_{1k}^2P_1(b_{1j},h_{1k})\frac{p_{1j}q_{1k}}{h_{1k}^2} 
	+0 \times P_1(0,0)[\beta_{\cBb\cHb} - \alpha_{\cBa\cHa}] \notag \\
	&\phantom{wwwww}
+\sum_{l=0}^{|\Gamma|-1}h_2^2(\gamma_l)P_2(b_2(\gamma_l),h_2(\gamma_l)) [\gamma_l-\gamma_{l-1}] \notag \\
&=\sum_{l=0}^{|\Gamma|-1}h_1^2(\gamma_l)P_1(b_1(\gamma_l),h_1(\gamma_l)) [\gamma_l-\gamma_{l-1}]
+\sum_{l=0}^{|\Gamma|-1}h_2^2(\gamma_l)P_2(b_2(\gamma_l),h_2(\gamma_l)) [\gamma_l-\gamma_{l-1}] \notag \\
&=\sum_{l=0}^{|\Gamma|-1} [h_1^2(\gamma_l)P_1(b_1(\gamma_l),h_1(\gamma_l))+h_2^2(\gamma_l)P_2(b_2(\gamma_l),h_2(\gamma_l))][\gamma_l-\gamma_{l-1}] \label{fineqn}.
\end{align}
Now an outage-free power allocation should satisfy 
\begin{align}
 h_1^2(\gamma_l)P_1(b_1(\gamma_l),h_1(\gamma_l))+h_2^2(\gamma_l)P_2(b_2(\gamma_l),h_2(\gamma_l)) \geq 2^{2\left(b_1(\gamma_{l})+b_2(\gamma_{l})\right)}-1 . \label{eq:lwrbnd:a}
\end{align}
Thus
\begin{align}
P_{avg}^{min}(1) \geq  \sum_{l=0}^{|\Gamma|-1}\left[2^{2\left(b_1(\gamma_{l})+b_2(\gamma_{l})\right)}-1\right][\gamma_l-\gamma_{l-1}] . \label{eq:lwrbnd:b}
\end{align}
But the RHS is indeed achieved by the power allocations in \eqref{eq:vf:pow:1} -- \eqref{eq:vf:pow:2}.
More specifically, \eqref{eq:eqlitythm} ensures equality in \eqref{eq:lwrbnd:a} for every $\gamma_l \in \Gamma$.

\end{IEEEproof}

\section{Proof of Lemma~\ref{lem:outage:2}} \label{sec:outage:2} 
The essential
ingredient for the proof is given in the lemma below.
\begin{lemma} \label{lem:recur:1}

Let $b_1,b_1^\prime, b_2, b_2^\prime$ be rates  such that $b_1^\prime \geq b_1$
and $b_2^\prime \geq b_2$, and let $h_1,h_1^\prime,h_2,h_2^\prime$ be arbitrary fading values. 
Let the power allocation functions $P_1(\cdot)$ and $P_2(\cdot)$ satisfy
$h_2^2P_2(b_2) + h_1^2P_1(b_1)  \geq 2^{2(b_1+b_2)}-1$ and $
 h_2^{\prime 2} P_2(b_2^\prime) + h_1^{\prime 2}P_1(b_1^\prime)  \geq 2^{2(b_1^\prime+b_2^\prime)}-1$.
If in addition $h_2^{\prime 2} P_2(b_2^\prime) + h_1^2P_1(b_1)  = 2^{2(b_1+b_2^\prime)}-1$, then
\begin{align*}
h_2^2 P_2(b_2) + h_1^{\prime 2} P_1(b_1^\prime) & \geq 2^{2(b_1^\prime+b_2)}-1.
\end{align*}
\end{lemma}

\begin{IEEEproof}
Observe that
\begin{align}
h_2^2 P_2(b_2) + h_1^{\prime 2} P_1(b_1^\prime) 
&= h_2^2P_2(b_2) + h_1^2P_1(b_1) + h_2^{\prime 2} P_2(b_2^\prime) + h_1^{\prime 2}P_1(b_1^\prime) 
- (h_2^{\prime 2} P_2(b_2^\prime) + h_1^2P_1(b_1)) \nonumber \\
& \geq 2^{2(b_1+b_2)} + 2^{2(b_1^\prime+b_2^\prime)} - 2^{2(b_1+b_2^\prime)} -1 \label{eq:le1}.
\end{align}
%
Note that
 $b_1+b_2 \leq b_1^\prime+b_2 \leq b_1^\prime+b_2^\prime$ and
 $b_1+b_2 \leq b_1+b_2^\prime \leq b_1^\prime+b_2^\prime$. Thus, 
\begin{align*}
2^{2(b_1^\prime+b_2^\prime)} + 2^{2(b_1+b_2)} \geq 2^{2(b_1+b_2^\prime)} + 2^{2(b_1^\prime+b_2)},
\end{align*}
by the convexity of $2^{2x}$ and Jensen's inequality. The lemma now follows from (\ref{eq:le1}).
\end{IEEEproof}
Let us now prove Lemma~\ref{lem:outage:2}.

\begin{IEEEproof}
 Consider any rate-channel pair $(b_{1j},h_{1k})$ and $(b_{2m},h_{2n})$ of user 1 and 2 respectively.
We will show that
\begin{align}
 h_{1k}^2P_1(b_{1j},h_{1k})+h_{2n}^2P_2(b_{2m},h_{2n}) \geq 2^{2(b_{1j}+b_{2m})}-1.
\end{align}
From the definition of $\gamma_l$, it follows that 
\begin{align}
 h_{1k}^2P_1(b_{1j},h_{1k})=h_1^2(\gamma_{l_1})P_1(b_1(\gamma_{l_1}),h_1(\gamma_{l_1})) \notag \\
 h_{2n}^2P_1(b_{1m},h_{1n})=h_2^2(\gamma_{l_2})P_2(b_2(\gamma_{l_2}),h_2(\gamma_{l_2})),\notag \\
\end{align}
for some $0 \leq l_1 \leq |\Gamma|-1,\, 0 \leq l_2 \leq |\Gamma|-1$.
So we need to prove that
\begin{align}
h_1^2(\gamma_{l_1})P_1(b_1(\gamma_{l_1}),h_1(\gamma_{l_1}))+ h_2^2(\gamma_{l_2})P_2(b_2(\gamma_{l_2}),h_2(\gamma_{l_2})) \geq 2^{2(b_1(\gamma_{l_1})+b_2(\gamma_{l_2}))}-1. \label{outcnd}
\end{align}
If $l_1=l_2$, then \eqref{outcnd} follows trivially from \eqref{eq:eqlitythm}. 
Assume without loss of generality that $l_1 > l_2$. 
The opposite case can be handled in a similar fashion.
Suppose it holds that 
\begin{align}
 h_1^2(\gamma_{l_1-1})P_1(b_1(\gamma_{l_1-1}),h_1(\gamma_{l_1-1}))
+h_2^2(\gamma_{(l_2)})P_2(b_2(\gamma_{(l_2)}),h_2(\gamma_{(l_2)})) \geq 2^{2(b_1(\gamma_{l_1-1})+b_2(\gamma_{(l_2)}))}-1.
\end{align}
Using this, along with  \eqref{eq:vf:pow:1} and \eqref{eq:eqlitythm} appropriately in  Lemma~\ref{lem:recur:1},
it follows that 
\begin{align}
h_1^2(\gamma_{l_1})P_1(b_1(\gamma_{l_1}),h_1(\gamma_{l_1}))+ h_2^2(\gamma_{l_2})P_2(b_2(\gamma_{l_2}),h_2(\gamma_{l_2})) \geq 2^{2(b_1(\gamma_{l_1})+b_2(\gamma_{l_2}))}-1. 
\end{align}
Thus by induction on $l_1$, \eqref{outcnd} holds for any $l_l > l_2$.
We next show that for $i=1,2$,
\begin{align}
 h_i^2(\gamma_{l})P_i(b_i(\gamma_{l}),h_i(\gamma_{l}))\geq 2^{2b_i(\gamma_{l})}-1.
\end{align}
We prove the case for $i=1$ by induction (the case of $i=2$ is similar).
The initial step in the induction is given by \eqref{eq:vf:pow:2}. Let
\begin{align}
 h_1^2(\gamma_{l-1})P_1(b_1(\gamma_{l-1}),h_1(\gamma_{l-1}))
	& \geq 2^{2b_1(\gamma_{l-1})}-1. \label{eqindstpin}
\end{align}
Then,
\begin{align}
 h_1^2(\gamma_{l})P_1(b_1(\gamma_{l}),h_1(\gamma_{l}))&= \left(h_1^2(\gamma_{l})P_1(b_1(\gamma_{l}),h_1(\gamma_{l}))
	+ h_2^2(\gamma_{l})P_2(b_2(\gamma_{l}),h_2(\gamma_{l}))\right) \notag\\
&\phantom{www}-\left(h_1^2(\gamma_{l-1})P_1(b_1(\gamma_{l-1}),h_1(\gamma_{l-1}))
	+ h_2^2(\gamma_{l})P_2(b_2(\gamma_{l}),h_2(\gamma_{l}))\right) \notag\\
&\phantom{www}+h_1^2(\gamma_{l-1})P_1(b_1(\gamma_{l-1}),h_1(\gamma_{l-1}))\notag \\
&\geq 2^{2(b_1(\gamma_{l})+b_2(\gamma_{l}))}-2^{2(b_1(\gamma_{l-1})+b_2(\gamma_{l}))}+2^{2(b_1(\gamma_{l-1})} -1\label{eqnsingout}\\
&=  2^{2(b_1(\gamma_{l})+b_2(\gamma_{l}))}-2^{2b_1(\gamma_{l-1})}(2^{2b_2(\gamma_{l})}-1)-1 \notag \\
&\geq  2^{2(b_1(\gamma_{l})+b_2(\gamma_{l}))}-2^{2b_1(\gamma_{l})}(2^{2b_2(\gamma_{l})}-1)-1 \label{eqnbgpkt}\\
&=  2^{2b_1(\gamma_{l})}-1. \notag
\end{align}
Here \eqref{eqnsingout} follows from \eqref{eq:vf:pow:1}, \eqref{eq:eqlitythm} and \eqref{eqindstpin}. Notice that \eqref{eqnbgpkt} follows from 
the fact that $b_1(\gamma_{l-1}) \leq b_1(\gamma_{l})$. This proves the result.
\end{IEEEproof}

 \section{Proof of Lemma~\ref{lem:conv:rate}}\label{sec:app1}
 \def\bp{b_1^{\prime}}
\def\bpp{b_1^{\prime \prime}}

The proof is similar to that in Lemma~\ref{lem:convex:sp}, we present it here for
completeness.

\begin{IEEEproof}
Consider three required packet-rates $\bp, b_1, \bpp$ at user~$1$ in 
the ascending order. W.l.o.g, take $b_1 = \lambda \bp + (1-\lambda) \bpp$ for some 
$\lambda \in (0,1)$. To prove the lemma, we will show that
$$
P_1(b_1) \leq \lambda P_1(\bp) + (1-\lambda)P_1(\bpp).
$$
By the power allocation in Lemma~\ref{lem:achieve:1}, we know
that for some $b_2 \in \mathcal B_2$, the rate-pair $(b_1,b_2)$ was assigned power
from the dominant face of a corresponding contra-polymatroid, i.e.
$$
\alpha_1 P_1(b_1) + \alpha_2 P_2(b_2) = 2^{2(b_1 + b_2)} - 1.
$$
We also know that for $\tilde b \in \{ \bp, \bpp\}$
\begin{align}
\alpha_1 P_1(\tilde b) + \alpha_2 P_2(b_2) &\geq  2^{2(\tilde b + b_2)} - 1 .
\end{align}
Taking a $\lambda$-linear combination, and using convexity,
\begin{align}
\alpha_1(\lambda P_1(\bp) + (1-\lambda) P_1(\bpp)) + \alpha_2 P_2(b_2) 
	&\geq \lambda 2^{2(\bp + b_2)}  + (1-\lambda) 2^{2(\bpp + b_2)}  - 1 \notag  \\
	&\geq 2^{2 (\lambda \bp + (1-\lambda)\bpp + b_2)} - 1  \notag \\
	&= 2^{2 (b_1 + b_2)} - 1  \notag \\
        &= \alpha_1 P_1(b_1) + \alpha_2 P_2(b_2).
\end{align} 
\end{IEEEproof}

\section{Proof of Proposition~\ref{prop:algo:conv}}\label{sec:prop:proof}

For the  BiSs $S_1$ and $S_2$, let $P_{s_1}(\cdot)$ and $P_{s_2}(\cdot)$ be the respective optimal 
power allocations obtained by Lemma~\ref{lem:achieve:1}.
By a slight abuse of notation, let us denote by  $P_{avg}(S_1, S_2)$ 
the average transmit sum-power  achieved by employing $(S_1,P_{s_1})$ and $(S_2,P_{s_2})$ respectively 
at the two transmitters. We first show that the average sum-power can be 
optimized by alternating the minimization of $P_{avg}(S_1,S_2)$ between
$S_1$ and $S_2$. On the other hand, though Algorithm~IterOpt alternates between $(S_1,S_2)$ and $(P_1,P_2)$, it 
still manages to  find the same minimum.
We start with the following lemma. 
\begin{lemma} \label{lem:strict:convex}
$P_{avg}(S_1,S_2)$ is strictly convex in $S_1$ for a given $S_2$.
\end{lemma}
\begin{IEEEproof}
Consider two possible BiS  schemes $S_a$ and $S_b$ for user~1, 
and let the second user employ the BiS  $S_2$. 
Let $(P_{1a},P_{2a})$ and $(P_{1b},P_{2b})$ denote the optimal power allocation schemes 
under the pair of schedulers $(S_a,S_2)$ and $(S_b,S_2)$ respectively. For $j \in \{a, b\}$,
the average
sum-power required at user~$l$ is denoted as $P_{lj}^{sum},  l\in\{1,2\}$.
%
Now, Lemma~\ref{lem:convex:sp} guarantees that
a $\lambda$-linear combination of $(S_a,S_2)$ and $(S_b,S_2)$ will be  an outage free
scheme. The average sum-power required for such a policy is 
$\lambda P_{1a}^{sum} + (1-\lambda)P_{1b}^{sum} + \lambda P_{2a}^{sum} + (1-\lambda)P_{2b}^{sum} $. 
It turns out that we can strictly improve this,
when $S_a$ and $S_b$ are not identical. Assume that there exists a rate $b_a$ ($b_b$) scheduled
in $S_a$ ($S_b$) such that
$b_{\lambda} =  \lambda b_a + (1-\lambda) b_b$ is scheduled at BiS~$S_1$, and  $b_a \neq b_b$.  
We now show that
the power allocation $P_{1\lambda}$ (linear combination of $P_{1a}$ and $P_{1b}$)  at user~$1$ 
is strictly sub-optimal. In particular, $P_{1\lambda}(\cdot)$ 
 fails to allocate  power for the rate $b_{\lambda}$ from the dominant face of any 
feasible contra-polymatroid. Thus, the power for $b_{\lambda}$ can be decreased without 
violating any other constraint or allocations.
To see this, for any $b_2$ scheduled at user~$2$, we have
\begin{align}
\alpha_1 P_{1\lambda}(b_{\lambda}) + \alpha_2 P_{2\lambda}(b_2) 
	&\geq \lambda 2^{2 (b_a + b_2)} + (1-\lambda) 2^{2(b_b + b_2)}  -1 \\
	&> 2^{2(\lambda b_a + (1-\lambda)b_b + b_2)} - 1. 
\end{align}  
The last inequality results from the strict convexity of $2^x, x \geq 0$. Thus, we can 
decrease $P_{1\lambda}(b_{\lambda})$ by a 
sufficiently small positive amount, and still guarantee the outage free nature of the scheme.
\end{IEEEproof}
Let us denote the minimal value of $P_{avg}(S_1,S_2)$ over $(S_1,S_2)$ as  $P_{AM}^*$. 
Consider an alternating minimization algorithm for minimizing $P_{avg}(S_1,S_2)$ over all
feasible distributed stationary schedulers. 
Lemma~\ref{lem:convex:sp} and Lemma~\ref{lem:strict:convex} ensures that
alternating the iterations between $S_1$ and $S_2$  will converge to the optimal value $P_{AM}^*$. 
This follows from the well known theory of
alternating minimization~\cite{Beck2014}, \cite{Grippo99}.
However, such an alternation among variables is not straight forward in our framework. In particular, 
the optimal power allocation in Lemma~\ref{lem:achieve:1} is jointly evaluated using the marginal
CDFs at the output 
of both the schedulers $S_1$ and $S_2$. While Algorithm~IterOpt circumvented this issue by alternating
over the variables $(S_1,S_2)$ and $(P_1,P_2)$,  fortunately, its terminal average power $P_{HALT}^*$ still
yields the correct minimum, i.e.
$$
P_{HALT}^* = P_{AM}^* = P_{avg}^{min}(D_{max}).
$$
To see this, let $C(P_1,P_2)$ denote the average sum-power for power policies $P_1$ and $P_2$
at the respective users. The associated schedulers will be clear from the context.
Assume that Algorithm~IterOpt terminates by converging to the BiS-CeN pairs
$(S_1^*,P_1^*)$ and $(S_2^*,P_2^*)$ for users $1$ and $2$ respectively. Observe that
$S_2^*$ is an optimal rate scheduler for the power control law $P_2^*$ (see Claim~\ref{claim:opt:su}).
In order to show that $(S_1^*,P_1^*)$ and $(S_2^*,P_2^*)$ are optimal,
let us now perform an alternate minimization between $(S_1,P_1)$ and $(S_2,P_2)$. 
For contradiction, assume that $(S_1^*,S_2^*)$ is not the optimal choice.
W.l.o.g, suppose we start with $(S_1^*,P_1^*)$ at the first user, 
and obtain another pair $(S_2^\prime,P_2^\prime)$ such that $P_2^* \neq
P_2^\prime$ and
\begin{align} \label{eq:algo:subopt}
 C(P_1^*,P_2^*) > C(P_1^*,P_2^\prime).
\end{align}
%
The inequality \eqref{eq:algo:subopt} suggests that
the point $(S_2^*,P_2^*)$ 
obtained via Algorithm~IterOpt was not the true optimum. Using $P_2^*$ and $P_2^\prime$,
let us construct another power function $P_2^o=\min (P_2^*,P_2^\prime)$. Clearly,
\begin{align*}
 C(P_1^*,P_2^\prime) > C(P_1^*,P_2^o).
\end{align*}
Notice that $(S_2^*,P_2^o)$ is also a feasible scheduler-power pair for user~$2$,
 and does not cause outage with any rate of user~$1$. The average sum-power under the
new power allocation
$(P_1^*,P_2^o)$ is strictly lower than that of either $(P_1^*,P_2^*)$ 
or $(P_1^*,P_2^\prime)$. However $P_2^*$ is an optimal power allocation function
for $S_2^*$. 
Hence
the  power-rate characteristics of  $P_2^*$ and $P_2^\prime$ must be identical. 
Once $P_2^*$ is fixed, $S_2^*$ is indeed an optimal scheduler by Claim~\ref{claim:opt:su}.
Thus $(S_1^*,S_2^*)$ is indeed the  stationary point of an alternating minimization
algorithm~\cite{Beck2014}, and in lieu of Lemma~\ref{lem:strict:convex} and Lemma~\ref{lem:convex:sp}, 
it achieves  the optimal value.

\section{Continuous Valued Packet Arrivals with Unit slot  Delay Constraint}\label{sec:cntvalpkts}

Consider packet arrivals with continuous valued rate requirements under a unit delay constraint.
Let the rate requirement be $B_i$ with respective CDFs $\phi_i(.)$ for user $i$. For
notational convenience, assume  the fading coefficients of user~$1$ and user~$2$ to be 
$1$ and $\sqrt{\alpha} $ respectively, with $\alpha \leq 1$. 
Define $\tilde\phi_1(x) :=1 - \alpha+ \alpha \phi_1(x)$ and $\tilde\phi_2(x) := \phi_2(x)$.
Let $\tilde b_i(y):= \tilde \phi_i^{-1}(y), i=1,2$,  as given in \eqref{eq:inv:cdf}.
The following power allocation minimizes the average sum  power for $D_{max}=1$.
\begin{theorem} \label{thm:conpack}
The power allocations
\begin{align}
 P_1(\tilde b_1(x))&=P_1(\tilde b_1(0))+2\int_{\tilde b_1(0)}^{\tilde b_1(x)}2^{2\left(y+\tilde\phi_2^{-1}\left(\tilde\phi_1(y)\right)\right)}dy\,, ~~~~~~0 \leq x \leq 1 \label{pwrminal1}\\
  P_2(\tilde b_2(x))&=P_2(\tilde b_2(0))+\frac{2}{\alpha}\int_{\tilde b_2(0)}^{\tilde b_2(x)}2^{2\left(y+\tilde\phi_1^{-1}\left(\tilde\phi_2(y)\right)\right)}dy\,, ~~~~~~0 \leq x \leq 1 \label{pwrminal2}
\end{align}
  for any $P_1(\tilde b_1(0)),~P_2(\tilde b_2(0))$ such that
\begin{align}
P_1(\tilde b_1(0)) &\geq 2^{2\tilde b_1(0)}-1 \\
\alpha P_2(\tilde b_2(0)) &\geq 2^{2 \tilde b_2(0)}-1 \\
P_1(\tilde b_1(0))+ \alpha P_2(\tilde b_2(0)) &= 2^{2(\tilde b_1(0)+\tilde b_2(0))}-1 
\end{align}
achieves $P_{avg}^{min}(1)$.

\end{theorem}

\begin{IEEEproof}
First, we prove a lower bound on the sum-power. Using the steps in \eqref{eq:ser:1} -- \eqref{eq:ser:3}, we get 
$$
\eE P_1(B_1) + \eE P_2(B_2) \geq \int_0^{1-\alpha} \frac{2^{2b_2(x)} - 1}{\alpha} dx 
	+ \int_{1-\alpha}^1 \frac{2^{2(b_1(\frac{x - 1 + \alpha}\alpha) + b_2(x))} - 1}{\alpha}.
$$
Since $\tilde b_1(x) = b_1(\frac{x-1 + \alpha}{\alpha})$ for $1-\alpha \leq x \leq 1$, we have
\begin{align}
P_{avg}^{min}(1) \geq \int_0^{1-\alpha} \frac{2^{2b_2(x)} - 1}{\alpha} dx 
	+ \int_{1-\alpha}^1 \frac{2^{2(\tilde b_1(x) + b_2(x))} - 1}{\alpha}. \label{eq:lwrsmbnd}
\end{align}

Next, we show that the power allocation given in \eqref{pwrminal1} and \eqref{pwrminal2} can achieve the lower bound.
For $y \leq 1-\alpha$, we have $\tilde  \phi_1^{-1}(\tilde \phi_2(y))=0$ and $\tilde b_1(y)=0$. Thus
$P_1(\tilde b_1(y))=0$, and 
\begin{align}
P_2(b_2(y)) = \frac{2^{2b_2(y)} - 1}{\alpha}
\label{eqpwrmin2}
\end{align}
For $1-\alpha \leq x \leq 1$,
\begin{multline}
 P_1(\tilde b_1(x))+\alpha P_2(\tilde b_2(x))  \\
	=P_1(\tilde b_1(0))+2\int_{\tilde b_1(0)}^{\tilde b_1(x)}2^{2\left(y+\tilde\phi_2^{-1}\left(\tilde\phi_1(y)\right)\right)}dy+\alpha P_2(\tilde b_2(0))+2\int_{\tilde b_2(0)}^{\tilde b_2(x)}2^{2\left(y+\tilde\phi_1^{-1}\left(\tilde\phi_2(y)\right)\right)}dy \notag.
\end{multline}
Substituting $\tilde\phi_2^{-1}\left(\tilde\phi_1(y)\right)=z$, we get
\begin{align}
 P_1(\tilde b_1(x))+\alpha P_2(\tilde b_2(x)) 
&=P_1(\tilde b_1(0))+\alpha P_2(\tilde b_2(0))+2\int_{\tilde b_2(0)}^{\tilde b_2(x)}2^{2\left(\tilde\phi_1^{-1}\left(\tilde\phi_2(z)\right)+z\right)}d(\tilde\phi_1^{-1}(\tilde\phi_2(z))) \notag \\
& \phantom{wwwww}+2\int_{\tilde b_2(0)}^{\tilde b_2(x)}2^{2\left(y+\tilde\phi_1^{-1}\left(\tilde\phi_2(y)\right)\right)}dy \notag \\
&=P_1(\tilde b_1(0))+\alpha P_2(\tilde b_2(0))+2\int_{\tilde b_2(0)}^{\tilde b_2(x)}2^{2\left(\tilde\phi_1^{-1}\left(\tilde\phi_2(z)\right)+z\right)}d(\tilde\phi_1^{-1}(\tilde\phi_2(z))+z) \notag \\
&=P_1(\tilde b_1(0))+\alpha P_2(\tilde b_2(0))+2^{2(\tilde b_1(x)+\tilde b_2(x))}-2^{2(\tilde b_1(0)+\tilde b_2(0))} \\
&=2^{2(\tilde b_1(x)+\tilde b_2(x))}-1 \label{eqpwrmin4}.
\end{align}
Now,  \eqref{eqpwrmin2} and  \eqref{eqpwrmin4} imply that we have equality in \eqref{eq:lwrsmbnd}.
 Note that from the above discussion, the sum power can be written as
\begin{align}
\eE P_1(B_1) + \eE P_2(B_2) = \int_0^1 \left( 2^{2(\tilde b_1(x)+\tilde b_2(x))}-1 \right) dx
\label{bndlowsm}.
\end{align}
Next we show that the power allocations in Theorem~\ref{thm:conpack} 
are outage free.
%
Substituting the lower limit of integration in $\tilde\phi_2^{-1}\left(\tilde\phi_1(y)\right)$ in 
\eqref{pwrminal1} and \eqref{pwrminal2}, we get
\begin{align*}
 P_1( b_1)=& P_1( \tilde b_1(0))+ 2\int_{\tilde b_1(0)}^{b_1}2^{2(y+\tilde b_2(0))}dy \\
&\geq 2^{2\tilde b_1(0)}-1+ 2\int_{\tilde b_1(0)}^{b_1}2^{2(y+\tilde b_2(0))}dy \\
&=2^{2\tilde b_1(0)}-1+2^{2\tilde b_2(0)}(2^{2b_1}- 2^{2\tilde b_1(0)}) \\
& \geq 2^{2b_1}-1.
\end{align*}
Similarly, we have $P_2(b_2)\geq \frac{1}{\alpha}(2^{2b_2}-1)$. Furthermore
\begin{multline}
P_1(b_1)+\alpha P_2(b_2)\\
=P_1( \tilde b_1(0))+\alpha P_2( \tilde b_2(0))+2\left[\int_{\tilde b_1(0)}^{b_1}2^{2\left(y+\tilde\phi_2^{-1}\left(\tilde\phi_1(y)\right)\right)}dy+\int_{\tilde b_2(0)}^{b_2}2^{2\left(y+\tilde\phi_1^{-1}\left(\tilde\phi_2(y)\right)\right)}dy\right] \notag .
\end{multline}
Substituting $\tilde\phi_1^{-1}(\tilde\phi_2(y))= z $ in the second integral above, we get

\begin{align*}
P_1(b_1)+\alpha P_2(b_2)
&=2^{2(\tilde b_1(0)+\tilde b_2(0))}-1+ 2\int_{\tilde b_1(0)}^{b_1}2^{2\left(y+\tilde\phi_2^{-1}\left(\tilde\phi_1(y)\right)\right)}dy\\
&\phantom{wwwww}+2\int_{\tilde b_1(0)}^{\tilde\phi_1^{-1}(\tilde\phi_2(b_2))}2^{2\left(z+\tilde\phi_2^{-1}\left(\tilde\phi_1(z)\right)\right)}d(\tilde\phi_2^{-1}(\tilde\phi_1(z)).
\end{align*}
Now, suppose $\tilde\phi_1^{-1}(\tilde\phi_2(b_2))\geq b_1$. Then
\begin{align}
P_1(b_1)+\alpha P_2(b_2) 
&=2^{2(\tilde b_1(0)+\tilde b_2(0))}-1+ 2\int_{\tilde b_1(0)}^{b_1}2^{2\left(y+\tilde\phi_2^{-1}\left(\tilde\phi_1(y)\right)\right)}d(y+\tilde\phi_2^{-1}(\tilde\phi_1(y))) \notag \\
&\phantom{wwwww}+2\int_{b_1}^{\tilde\phi_1^{-1}(\tilde\phi_2(b_2))}2^{2\left(z+\tilde\phi_2^{-1}\left(\tilde\phi_1(z)\right)\right)}d(\tilde\phi_2^{-1}(\tilde\phi_1(z))) \notag \\
& = 2^{2(\tilde b_1(0)+\tilde b_2(0))}-1+ 2^{2(b_1+\tilde\phi_2^{-1}(\tilde\phi_1(b_1)))}-2^{2(\tilde b_1(0)+\tilde b_2(0))} \notag\\
&\phantom{wwwww}+2\int_{b_1}^{\tilde\phi_1^{-1}(\tilde\phi_2(b_2))}2^{2\left(z+\tilde\phi_2^{-1}\left(\tilde\phi_1(z)\right)\right)}d(\tilde\phi_2^{-1}(\tilde\phi_1(z))) \label{eqsub2}\\
&\geq  2^{2(\tilde b_1(0)+\tilde b_2(0))}-1+ 2^{2(b_1+\tilde\phi_2^{-1}(\tilde\phi_1(b_1)))}-2^{2(\tilde b_1(0)+\tilde b_2(0))} \notag \\
&\phantom{wwwww}+2\int_{b_1}^{\tilde\phi_1^{-1}(\tilde\phi_2(b_2))}2^{2\left(b_1+\tilde\phi_2^{-1}\left(\tilde\phi_1(z)\right)\right)}d(\tilde\phi_2^{-1}(\tilde\phi_1(z))) \label{eqsubl} \\
& \geq 2^{2(b_1+b_2)}-1.
\end{align}
The inequality in \eqref{eqsubl} was obtained by substituting  the lower bound of $z$ in \eqref{eqsub2}.
The other case when $\tilde\phi_1^{-1}(\tilde\phi_2(b_2))< b_1$ can be handled in a similar fashion. 
Thus the given power allocations are outage free, which proves the theorem.

\end{IEEEproof}

\end{appendices}

\bibliographystyle{IEEEtran}
\bibliography{../../../biblio/poster}{}


\end{document}